\documentclass[twocolumn,trackchanges]{aastex7}

\usepackage{graphicx}    
\usepackage{subcaption}  
\usepackage{placeins}  %

\begin{document}

\title{Mass$-$Orbital Period Distribution of Massive White Dwarfs Formed through Stable Mass Transfer}

\author[orcid=0009-0001-6159-2236]{Rizhong Zheng}
\affiliation{Yunnan Observatories, Chinese Academy of Sciences (CAS), Kunming 650216, People's Republic of China}
\affiliation{International Centre of Supernovae (ICESUN), Yunnan Key Laboratory of Supernova Research, Kunming 650216, People's Republic of China}
\affiliation{University of Chinese Academy of Sciences, Beijing 100049, People's Republic of China}
\email{zhengrizhong23@mails.ucas.ac.cn}

\author[orcid=0000-0002-6398-0195]{Hongwei Ge\textsuperscript{*}}
\affiliation{Yunnan Observatories, Chinese Academy of Sciences (CAS), Kunming 650216, People's Republic of China}
\affiliation{International Centre of Supernovae (ICESUN), Yunnan Key Laboratory of Supernova Research, Kunming 650216, People's Republic of China}
\affiliation{University of Chinese Academy of Sciences, Beijing 100049, People's Republic of China}
\email{gehw@ynao.ac.cn}

\author[0000-0002-1556-9449]{Christopher A. Tout\textsuperscript{*}}
\affiliation{Institute of Astronomy, The Observatories, University of Cambridge, Madingley Road, Cambridge, CB3 0HA, UK}
\email{cat@ast.cam.ac.uk}

\author[0009-0006-9211-2860]{Hailiang Chen} 
\affiliation{Yunnan Observatories, Chinese Academy of Sciences (CAS), Kunming 650216, People's Republic of China}
\affiliation{International Centre of Supernovae (ICESUN), Yunnan Key Laboratory of Supernova Research, Kunming 650216, People's Republic of China}
\affiliation{University of Chinese Academy of Sciences, Beijing 100049, People's Republic of China}
\email{chenhl@ynao.ac.cn}

\author[0000-0002-1421-4427]{Zhenwei Li} 
\affiliation{Yunnan Observatories, Chinese Academy of Sciences (CAS), Kunming 650216, People's Republic of China}
\affiliation{International Centre of Supernovae (ICESUN), Yunnan Key Laboratory of Supernova Research, Kunming 650216, People's Republic of China}
\affiliation{University of Chinese Academy of Sciences, Beijing 100049, People's Republic of China}
\email{lizw@ynao.ac.cn}

\author[0000-0003-4265-7783]{Dengkai Jiang} 
\affiliation{Yunnan Observatories, Chinese Academy of Sciences (CAS), Kunming 650216, People's Republic of China}
\affiliation{International Centre of Supernovae (ICESUN), Yunnan Key Laboratory of Supernova Research, Kunming 650216, People's Republic of China}
\affiliation{University of Chinese Academy of Sciences, Beijing 100049, People's Republic of China}
\email{dengkai@ynao.ac.cn}

\author[0000-0002-3084-5157]{Chengyuan Li}
\affiliation{School of Physics and Astronomy, Sun Yat-sen University, Zhuhai 519082, People’s Republic of China}
\email{lichengy5@mail.sysu.edu.cn}

\author[0000-0003-0220-7112]{Zhijia Tian}
\affiliation{Department of Astronomy, Key Laboratory of Astroparticle Physics of Yunnan Province, Yunnan University, Kunming 650200, People’s Republic of China}
\email{tianzhijia@ynu.edu.cn}

\author[0000-0002-0378-2023]{Bo Ma}
\affiliation{School of Physics and Astronomy, Sun Yat-sen University, Zhuhai 519082, People’s Republic of China}
\email{mabo8@mail.sysu.edu.cn}

\author[0009-0001-3638-3133]{Lifu Zhang}
\affiliation{Yunnan Observatories, Chinese Academy of Sciences (CAS), Kunming 650216, People's Republic of China}
\affiliation{International Centre of Supernovae (ICESUN), Yunnan Key Laboratory of Supernova Research, Kunming 650216, People's Republic of China}
\affiliation{University of Chinese Academy of Sciences, Beijing 100049, People's Republic of China}
\email{zhanglifu@ynao.ac.cn}

\author[0009-0001-4332-6575]{Jian Mou}
\affiliation{Yunnan Observatories, Chinese Academy of Sciences (CAS), Kunming 650216, People's Republic of China}
\affiliation{International Centre of Supernovae (ICESUN), Yunnan Key Laboratory of Supernova Research, Kunming 650216, People's Republic of China}
\affiliation{University of Chinese Academy of Sciences, Beijing 100049, People's Republic of China}
\email{moujian@ynao.ac.cn}

\author[0000-0001-5284-8001]{Xuefei Chen\textsuperscript{*}}
\affiliation{Yunnan Observatories, Chinese Academy of Sciences (CAS), Kunming 650216, People's Republic of China}
\affiliation{International Centre of Supernovae (ICESUN), Yunnan Key Laboratory of Supernova Research, Kunming 650216, People's Republic of China}
\affiliation{University of Chinese Academy of Sciences, Beijing 100049, People's Republic of China}

\email{cxf@ynao.ac.cn}

\author[0000-0001-9204-7778]{Zhanwen Han\textsuperscript{*}}
\affiliation{Yunnan Observatories, Chinese Academy of Sciences (CAS), Kunming 650216, People's Republic of China}
\affiliation{International Centre of Supernovae (ICESUN), Yunnan Key Laboratory of Supernova Research, Kunming 650216, People's Republic of China}
\affiliation{University of Chinese Academy of Sciences, Beijing 100049, People's Republic of China}
\email{zhanwenhan@ynao.ac.cn}

\renewcommand{\thefootnote}{}
\footnotetext{*Corresponding authors: Hongwei Ge, gehw@ynao.ac.cn;
	Christopher A. Tout, cat@ast.cam.ac.uk;
	Xuefei Chen, cxf@ynao.ac.cn;
	Zhanwen Han, zhanwenhan@ynao.ac.cn.}

\renewcommand{\thefootnote}{\arabic{footnote}}
\begin{abstract}
White dwarfs (WDs) in binaries can form through either the stable mass-transfer process or common envelope evolution (CEE). Compared to CEE, the stable mass-transfer process can lead to a distinct mass$-$orbital period ($M_{ \mathrm{WD}}$--$P_{ \mathrm{orb}}$) relation. Thus, this relation of WDs contains the information about the evolution channels. We can study the relation in WD binary systems to determine whether their progenitors undergo a CEE. We use the stellar evolution code MESA as our primary computational tool and adopt the quasi-adiabatic criterion to ensure that our models satisfy the conditions for stable mass transfer. Our study considers different mass-transfer schemes, varying metallicities, and the relation for both low-mass and intermediate-mass progenitors. Previous studies have focused on the relation for low-mass progenitors, which cannot explain some long-period, high-mass WD binaries. Our results show that the $M_{ \mathrm{WD}}$--$P_{ \mathrm{orb}}$ distribution for intermediate-mass progenitors whose cores remain nondegenerate prior to central helium burning can account for the formation channels of long-period and massive WD binaries.
\end{abstract}


\section{Introduction}

White dwarfs (WDs) are the final evolutionary state of the majority of
stars and typically have masses between about 0.15 and $1.4 \, M_\odot$,
with a distribution that peaks around $0.6\, M_\odot$ \citep{saumonCurrentChallengesPhysics2022}.  In binary systems, their properties powerfully constrain
stellar and binary evolution, including mass transfer, angular
momentum loss and compact-object formation  (e.g., \citealt{palaWhiteDwarfBinaries2025}).

A key aspect of WD formation in binaries is the evolutionary channel
through which the system passes.  Two principal channels are generally
considered: stable mass transfer and common envelope-evolution (CEE;
\citealt{1976IAUS...73...75P}).  These channels lead to markedly different orbital
configurations.  In particular, CEE typically produces compact systems
\citep{ivanovaCommonEnvelopeEvolution2013}, whereas stable mass transfer can result in
significantly wider orbits  \citep{paczynskiEvolutionaryProcessesClose1971, hurleyEvolutionBinaryStars2002}.
Distinguishing between these channels is therefore a central problem
in binary stellar evolution \citep{geAdiabaticMassLoss2010, geCommonEnvelopeEvolution2024}.

One useful diagnostic is the relation between the WD mass and the
binary orbital period.  For systems formed
via stable mass transfer, this relation can be derived from the
connection between the donor's core mass and radius, combined with
Roche-lobe geometry and Kepler's laws \citep{paczynskiEvolutionaryProcessesClose1971,rappaportRelationWhiteDwarf1995}.  Previous studies have established such relations
for low-mass progenitors, in particular for systems undergoing mass
transfer during the red-giant phase \citep{thomasmFormationMillisecondPulsars1999,linLMXBIMXBEVOLUTION2011,smedleyNatureMillisecondPulsars2014}.  These models successfully explain
many observed systems.

However, an increasing number of observed binaries do not follow the
canonical $M_{\rm WD}$--$P_{\rm orb}$ relation.  For example,
long-period systems such as B0820+02 \citep{kawaharaDiscoveryThreeSelflensing2018}, as well
as several recently identified self-lensing binaries (e.g., KIC
03835482, KIC 06233093, and KIC 12254688, \citealt{kawaharaDiscoveryThreeSelflensing2018}), lie
significantly below the predicted relation.  In addition, a large
population of long-period WD binaries identified in Gaia DR3 also
show similar deviations \citep{hallakounDeficitMassiveWhite2024,shahafTriageGaiaDR32024}.  Their long orbital
periods strongly disfavor a CEE origin, suggesting that current models
of stable mass transfer may be incomplete.

These discrepancies indicate that the standard $M_{\rm WD}$--$P_{\rm
	orb}$ relation, derived primarily for low-mass progenitors with
degenerate cores, may not be applicable to all systems.  In particular,
binaries originating from intermediate-mass progenitors with
nondegenerate cores before central helium burning may follow a different evolutionary pathway,
leading to distinct mass–period relations.

In this work, we aim to resolve these discrepancies by systematically
investigating the $M_{\rm WD}$--$P_{\rm orb}$ relation for both low-
and intermediate-mass progenitors under stable mass transfer. Using
detailed binary evolution calculations, we explore how the relation
depends on progenitor mass, metallicity and mass-transfer
prescriptions.  Our goal is to determine whether the observed outliers
can be explained within the framework of stable mass transfer, without
invoking CEE.

In Section~\ref{sec:method}, we describe our numerical methods and physical
assumptions.  In Section~\ref{sec:resault}, we present the resulting $M_{\rm WD}$--$P_{\rm orb}$ distribution of low and massive WDs\footnote{Previous research has largely centered on WDs with $ M \lesssim 0.5 \,  M_{\odot}$. Here, we use 'massive WDs' to denote systems with $M_{\text{WD}} > 0.5 \, M_{\odot}$} and compare them with observations.  In Section~\ref{sec:discussion}, we examine the influence of the accretor mass and accretion efficiency, and discuss the limitations of the current models at extreme mass ratios. In Section~\ref{conclusion}, we summarize our conclusions.

\section{Method} \label{sec:method}
To investigate the distribution of WD masses and orbital
periods produced under different conditions,
we perform detailed binary evolution calculations of systems
undergoing stable mass transfer from low- and intermediate-mass
progenitors.
We use the stellar evolution code MESA \citep{paxtonMODULESEXPERIMENTSSTELLAR2011,paxtonMODULESEXPERIMENTSSTELLAR2013,paxtonMODULESEXPERIMENTSSTELLAR2015,paxtonModulesExperimentsStellar2018,paxtonModulesExperimentsStellar2019,jermynModulesExperimentsStellar2023}, version 24.08.1, as our computational tool. The accretors are set to point masses. 

We adopt two specific values for the accretor mass ($M_{\rm a}$): $1.4 \, M_\odot$ and $2.3 \, M_\odot$. These values are chosen to represent a broad range of potential companions observed in binary systems. The $1.4 \, M_\odot$ case corresponds to both the canonical neutron star (NS) mass and the typical mass of main-sequence (MS) companions in systems such as KICs 03835482, 06233093, and 12254688 \citep{kawaharaDiscoveryThreeSelflensing2018}. The $2.3 \, M_\odot$ case represents the upper limit of observed NS masses \citep{bassaLOFARDiscoveryFastestspinning2017, romaniPSRJ09520607Fastest2022}. Our model grid covers a wide range of initial orbital periods, from 3 to 1950 days, enabling a comprehensive comparison at the different stages of mass transfer onset.

The donors evolve from the zero-age main sequence (ZAMS) to the WD cooling sequence. The donors' masses ($M_{\rm d}$) range from $1.0 \, M_\odot$ to $4.0 \, M_\odot$, which is motivated by our test calculations showing that systems with a $1.4\,M_\odot$ accretor become largely unstable above $\sim 2.3\,M_\odot$, while for a $2.3\,M_\odot$ accretor stable evolution can extend up to $\sim 4.0\,M_\odot$ for solar and one-tenth solar metallicities considered. During the RGB phase, we use the Reimers wind prescription \citep{reimers1975} with a scale factor of $\eta _\mathrm{r} =0.25$ \citep{miglioAgeDissectionMilky2021}. While during the AGB phase, the stellar wind is set to the Blocker scheme \citep{bloeckerStellarEvolutionLow1995} with a scale factor of $\eta _\mathrm{b}=0.05$ for the donors with masses below $ 3.0 \, M_\odot$, and $\eta _\mathrm{b}=0.1$ for the donors with masses above $ 3.0 \, M_\odot$, following the typically adopted values in
\citet{paxtonMODULESEXPERIMENTSSTELLAR2011}.
This choice reflects the expectation that the
mass-loss efficiency of AGB stars depends on their stellar
properties, in particular their mass
\citep{hofnerMassLossStars2018}.
Our mixing length parameter is set to 1.76 with reference to \citealt{trampedachImprovementsStellarStructure2014}. 
We use the exponential convective overshooting scheme and include both core and shell overshooting. The overshooting parameter $f$ is adopted from the calibration of \citet{choiMESAISOCHRONESSTELLAR2016}, with $f=0.0174$ for shell overshooting and $f=0.016$ for core overshooting. The corresponding $f_0$ values are set to $0.0087$ and $0.08$ for shell and core overshooting, respectively, following the publicly available MESA inlist provided by \citet{temminkCopingLossStability2023}. 
We regard $\mathrm{Z}\, =\, 0.0134$ as the solar metallicity \citep{asplundChemicalCompositionSun2009}, and consider metallicities of $Z_\odot$, $0.1 \, Z_\odot$, $10^{-3} \, Z_\odot$, and $10^{-4} \, Z_\odot$ in our models.

We adopt the default prescriptions in MESA for angular momentum loss, encompassing contributions from gravitational-wave radiation \citep[e.g.,][]{petersGravitationalRadiationMotion1964}, magnetic braking \citep[e.g.,][]{rappaportNewTechniqueCalculations1983}, and mass loss \citep[e.g.,][]{tauris2006formation}, as implemented in MESA \citep{paxtonMODULESEXPERIMENTSSTELLAR2015}. For long-period binaries, angular momentum loss from mass loss will be the dominant factor. Mass loss in binaries can be attributed to stellar winds and mass transfer.

During mass transfer, both NS and MS accretors may lose part of the transferred material in the vicinity of the accretor through several channels. In our parameterized scheme, a fraction of this material is assumed to be lost locally and is described by the MESA parameter $\beta$.

For compact accretors, the prescription of mass loss around accretors often referred to as isotropic re-emission, which may occur through mechanisms 
such as jets or disk winds \citep{tauris2023physics}. Other mass-loss 
channels, for example outflows through the outer Lagrange points $\rm L_2$ (e.g., \citealt{linialMasslossL2Lagrange2017}), may also 
be possible in certain situations. For NS accretors, we set $\beta = 0.7 $, which assumes that $70 \, \%$ of their mass is lost around the vicinity of the accretor \citep{antoniadisRelativisticPulsarwhiteDwarf2012}. 
In the process of mass transfer, the corrections of mass-transfer rates from the Eddington limit and X-rays \citep{podsiadlowskiFormationEvolutionBlack2003} are taken into account.

Currently, the precise physical mechanism of jets remains unclear. Besides, there is limited research on disk winds, and the accretion rates at different disk positions may vary \citep{kosecVerticalWindStructure2023}. When the materials are transferred to the compact accretor at super-Eddington rates, the radiation pressure generated by the X-rays near the surface of the compact accretor will remove some of the material. Even if the mass-transfer rate is higher than the Eddington limit before it goes through disk winds, the mass-transfer rate can decline to below the Eddington limit near the surface of the accretor because the disk winds may blow away some of the transferred mass. 
In our calculation of mass loss caused by mass transfer, we first calculate the impact of $\beta$, then $(1-\beta) \, \dot{M}_{\mathrm{d}}$ will be restricted by the Eddington limit.
  
 There will still be a loss of a fraction of the transferred mass, even if the $(1-\beta) \, \dot{M}_{\mathrm{d}}$ is below the Eddington limit. 
In the mass-transfer process, a fraction $\eta$ of the transferred mass is lost from the accretor and converted into radiation, which escapes as X-rays, producing the observed accretion luminosity:
\begin{equation}
    L_{\mathrm{acc}}=\eta\,(1-\beta) \,\dot{M}_{\mathrm{d}} \,c^2,  
\end{equation}
where $c$ denotes the light speed. We use $\eta=0.15$ for the accretors  \citep{lattimerNeutronStarObservations2007}. Taking into account both $\beta$ and $\eta$, this means that around  $75 \%$ mass is lost for NS accretors when the mass transfer rates are below the Eddington limit, which is considered reasonable \citep{antoniadisRelativisticPulsarwhiteDwarf2012}. .
The calculation procedure is
\begin{equation}
	\dot{M}_{\mathrm{a}} = \text{min}\,[ \dot{M}_{\mathrm{Edd}}, (1-\beta) \dot{M}_{\mathrm{d}}]\,(1-\eta).
\end{equation}
When the mass-transfer rates overrun the Eddington limit, they will be corrected by both the Eddington limit and X-ray radiation:
\begin{equation}
	\dot{M}_{\mathrm{a}} =(1- \eta\,) \dot{M}_{\mathrm{Edd}}.
\end{equation}

While the aforementioned prescriptions ($\beta = 0.7$ and X-ray corrections) are tailored for compact accretors such as neutron stars, we also consider a case with $\beta = 0.5$ (e.g., \citealt{shaoPopulationSynthesisGalactic2021}) to facilitate comparisons with observed WD+MS binaries. Previous studies have demonstrated that at the low-mass end, the mass-transfer efficiency has a negligible impact on the final $M_{\rm WD}$--$P_{\rm orb}$ relation; hence, we primarily apply the $\beta = 0.5$ setting to our intermediate-mass models. In these calculations, we disable the Eddington limit and radiation-corrected transfer rates, as the MS accretors do not experience the extreme radiation pressure characteristic of compact objects.

 To ensure our models are stable during the mass-transfer process, we apply the quasi-adiabatic criterion of \citet{temminkCopingLossStability2023} as our stability criterion. 
 
 Even if the timescale of mass transfer is shorter than the global thermal timescale, the mass-transfer process can still be stable, attributed to surface layers' thermal readjustment. We calculate the local thermal timescale as a function of mass coordinate \citep{kippenhahn2012},
 \begin{equation}
 	\tau_{\text{th}}(m) = \frac{1}{L} \int_{m}^{M} c_P(\widetilde{m})T(\widetilde{m})d\widetilde{m},
 \end{equation}
where $M$ denotes the coordinate of the outermost layer. $\dot{M}_{\text{th}}(m)$ represents the minimum mass-transfer rate at which the star cannot readjust itself as its mass decreases from $M$ to $m$. We can calculate it as \citep{temminkCopingLossStability2023}
\begin{equation}
	\dot{M}_{\text{th}}(m) = - \frac{M - m}{\tau_{\text{th}}(m)}.
\end{equation}

The local thermal-timescale mass-transfer rate,
$\dot{M}_{\mathrm{th}}(m)$, depends on the mass coordinate $m$
within the donor star and therefore varies throughout the stellar
interior. To characterize the overall thermal response of the donor to mass loss, 
the critical mass-transfer rate is defined as the maximum of the local 
thermal-timescale mass-transfer rate over all mass shells:
\begin{equation}
\label{eq:crititcal rates}
	\dot{M}_{\text{th,crit}} = \max \, \dot{M}_{\text{th}}(m).
\end{equation}

\begin{figure}[htbp] 
\centering \includegraphics[width=0.9\columnwidth]{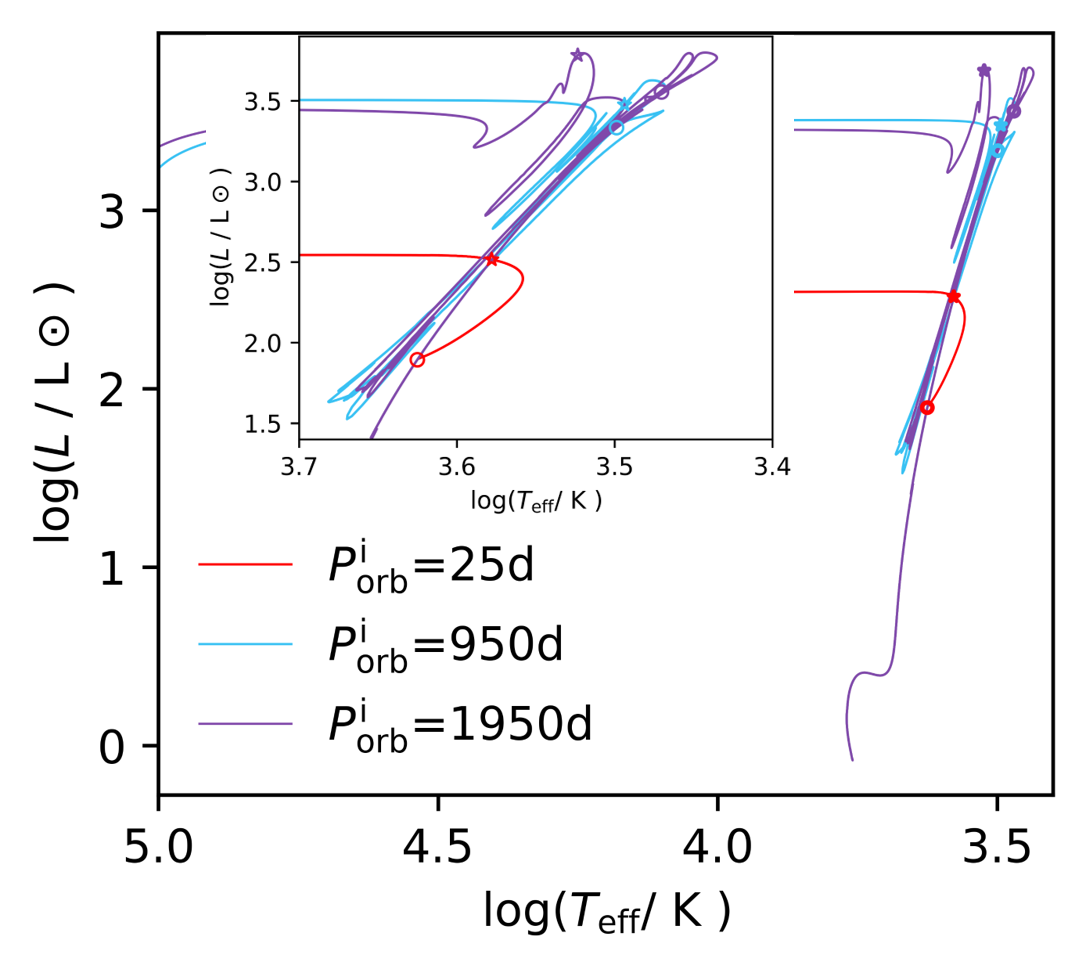} 
\caption{Evolution in the H-R diagram of the donor star for binary models with an initial donor mass of $1.0 \, M_\odot$, an initial accretor mass of $1.4 \, M_\odot$, and an initial metallicity of $Z_\odot$. The tracks illustrate the complete evolutionary path for various initial orbital periods, starting from the ZAMS through the mass-transfer phases and ending at the WD cooling stage. Circles and star symbols denote the onset and termination of the mass-transfer phase for each model, respectively. The upper-left panel provides a zoomed-in view of the evolutionary tracks during the mass-transfer stage.} 
\label{fig1} 
\end{figure}

\begin{figure}[htbp] 
\centering \includegraphics[width=0.9\columnwidth]{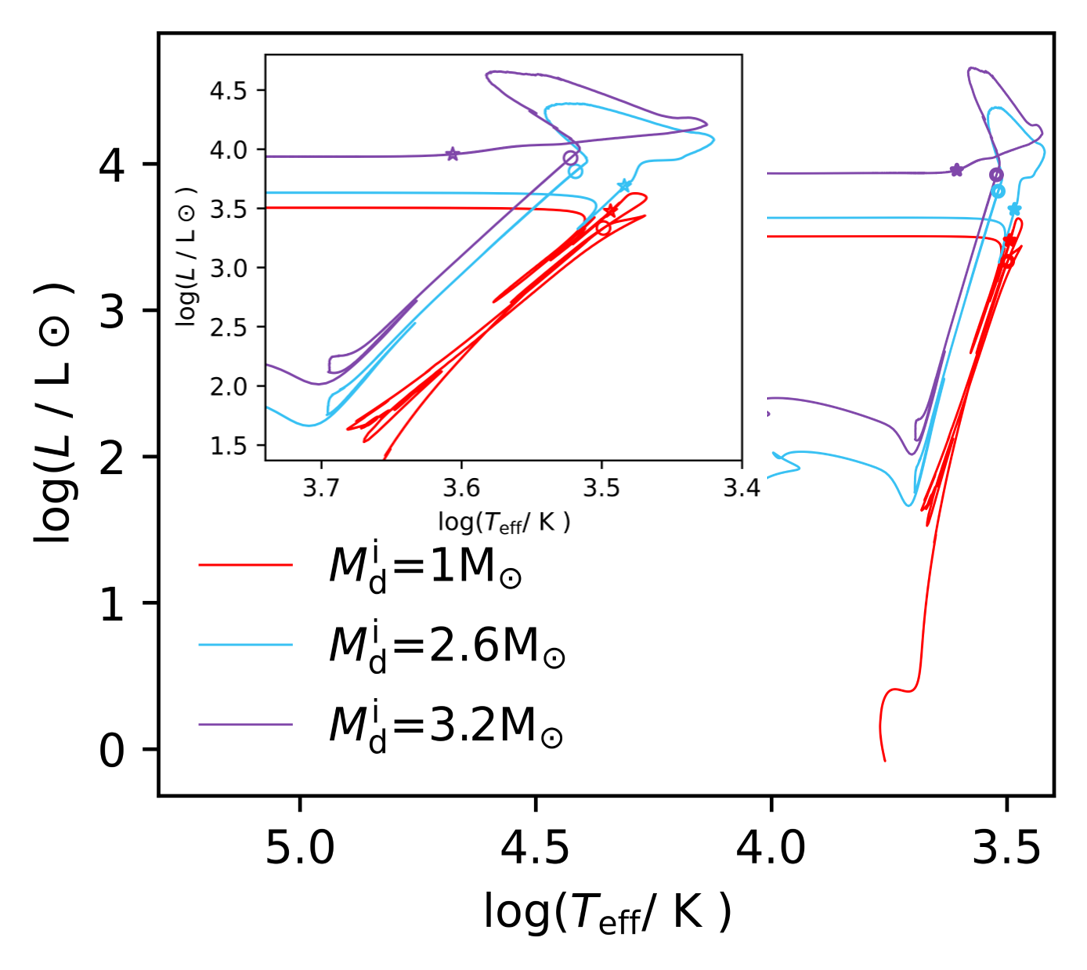} 
\caption{Similar to Figure ~\ref{fig1}, but for models with different initial donor masses ($M_{\mathrm{d}}^{\mathrm{i}}$) and the same initial orbital period ($P_{\mathrm{orb}}^{\mathrm{i}}$) of 950 days.}
\label{fig-dm} 
\end{figure}

\section{Result} 
\label{sec:resault}

Figure~\ref{fig1} shows the Hertzsprung–Russell (HR) diagrams for three models that share the same initial donor mass and solar metallicity, but differ in their initial orbital periods. The circles denote the onset of  mass transfer, while the star symbols mark the end of the  mass transfer.
We define the onset and termination of mass transfer as the points where the mass-transfer rate through the inner Lagrangian point, $\log \dot{M}_{\mathrm{d}}$, rises above and falls below $-9$, respectively. \footnote{Unless otherwise stated, $\log$ denotes the base-10 logarithm, and all mass-transfer rates in this paper are expressed in units of $M_\odot\,\rm {yr^{-1}}$.}

\begin{figure*}[htbp]
    \centering
\begin{subfigure}[b]{0.45\textwidth}
        \centering
        \includegraphics[width=\textwidth]{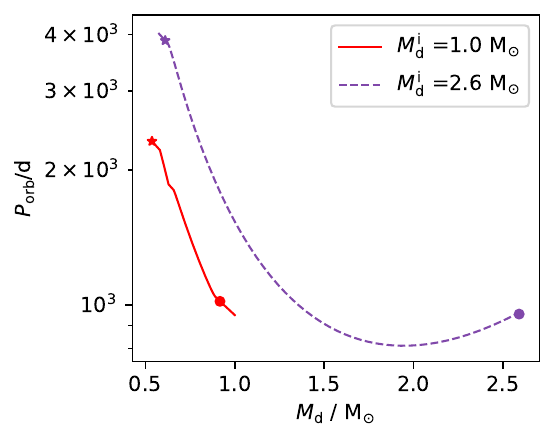}
        \caption{}
        \label{fig:sub1}
\end{subfigure}
    \hfill
\begin{subfigure}[b]{0.45\textwidth}
        \centering
        \includegraphics[width=\textwidth]{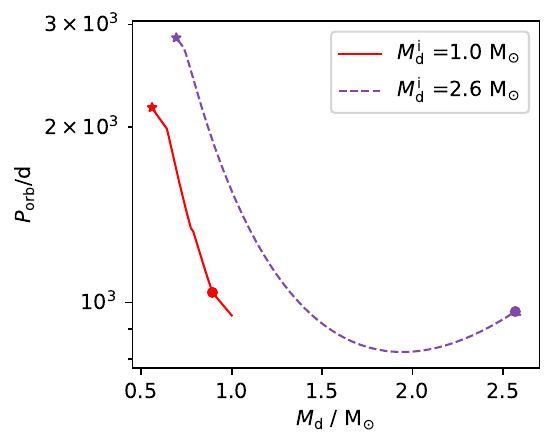}
        \caption{}
        \label{fig:sub2}
\end{subfigure}

\caption{Evolution of the orbital period as a function of decreasing donor mass for binary models with an initial orbital period of 950 days and an initial accretor mass of $1.4 \, M_\odot$. The two panels compare the evolution under different initial metallicities: (a) solar metallicity ($Z^{\mathrm{i}} = Z_\odot$) and (b) subsolar metallicity ($Z^{\mathrm{i}} = 0.1 \, Z_\odot$). In each panel, tracks for models with different initial donor masses ($1.0 \, M_\odot$ and $2.6 \, M_\odot$) are shown.
The onset and termination of the mass-transfer phase are denoted by solid circles and star symbols, respectively.
}
\label{fig:both}
\end{figure*}

\begin{figure}[htbp] 
	\centering \includegraphics[width=0.9\columnwidth]{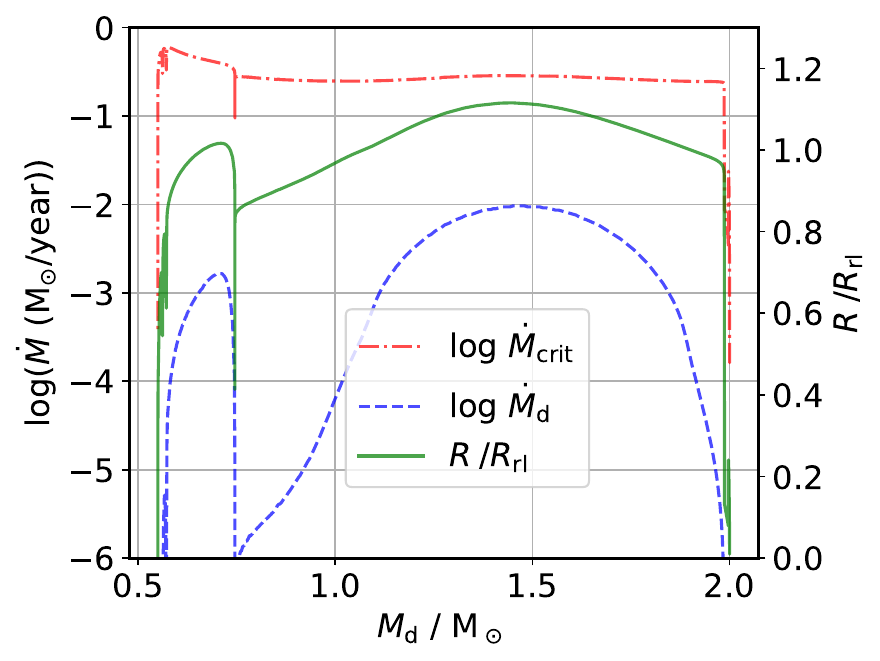} 
	\caption{Comparison of the donor mass-transfer rate ($\dot{M}_{\mathrm{d}}$) with the critical mass-transfer rate ($\dot{M}_{\mathrm{crit}}$, calculated via Eq.~\ref{eq:crititcal rates}) as a function of donor mass for a representative model undergoing stable mass transfer. The model parameters are: initial donor mass $M_{\mathrm{d}}^{\mathrm{i}} = 2.0 \, M_\odot$, initial accretor mass $M_{\mathrm{a}}^{\mathrm{i}} = 1.4 \, M_\odot$, initial orbital period $P_{\mathrm{orb}}^{\mathrm{i}} = 303$ days, and initial metallicity $Z^{\mathrm{i}} = 0.1 \, Z_\odot$. The red dashed-dotted line and blue dashed line (left axis) represent $\log \dot{M}_{\mathrm{crit}}$ and $\log \dot{M}_{\mathrm{d}}$, respectively. Additionally, the ratio of the donor radius to its Roche-lobe radius ($R\,/R_{\mathrm{rl}}$; green solid line, right axis) is shown for the same model.  This model undergoes stable mass transfer.
}
	\label{fig:3-2-stable} 
\end{figure}

Notably, longer initial orbital periods suggest that donors will overfill their Roche lobes and begin to transfer mass at later evolutionary stages, sometimes during the AGB phase. In Figure~\ref{fig1}, the model with an initial orbital period of 25 days underwent mass transfer in Case B, while the model with an initial orbital period of 950 days underwent mass transfer in Case BC (systems in which mass transfer starts during Case B and resumes during the AGB phase; see \citealt{tauris2023physics}), with most of the mass being transferred during the AGB phase. Furthermore, the model with the initial orbital period of 1950 days underwent mass transfer in Case C. 
The chosen range of initial orbital periods is designed so that a significant fraction of models with different donor masses and metallicities can undergo mass transfer during the AGB phase.

Figure~\ref{fig-dm} presents the HR diagrams for the three models with identical initial parameters except for the initial donor mass. Compared to the model of initial mass $1\,M_\odot$, the models with initial masses of $2.6\,M_\odot$ and $3.2\,M_\odot$ underwent mass transfer entirely during the AGB phase, corresponding to Case C, even though they shared the same initial orbital period. Models with larger initial masses are more likely to undergo mass transfer during the AGB phase with the same initial orbital period.

Binary systems with larger $q$ values imply a higher probability of undergoing unstable mass transfer. Based on the quasi-adiabatic criterion \citep{temminkCopingLossStability2023}, donors with larger donor radius when they transfer their mass are generally more likely to have higher critical mass-transfer rates. A shorter initial period typically results in a smaller donor radius at the onset of mass transfer, thereby leading to a lower critical mass-transfer rate. Our models with a larger mass ratio sometimes will enter unstable mass-transfer processes with relatively shorter initial periods for this reason.

The orbital period evolves in response to mass transfer and the
associated redistribution of orbital angular momentum.
In particular, when mass is transferred from a less massive donor
to a more massive accretor, the orbit tends to widen, leading to
an increase in the orbital period. This behavior can be seen in
our evolutionary sequences shown in Figure~\ref{fig:both}.

Before presenting the results, we note that, in all models discussed in Sections~\ref{Mass-Transfer Instability}–\ref{sec-Distribution for Progenitors with intermediate mass}, the accretor is assumed to be a 1.4 $M_\odot$ neutron star. For these models, the mass-transfer efficiency parameter is set to $\beta = 0.7$, with additional limitations imposed by the Eddington accretion rate and radiation feedback.

\subsection{Mass-transfer Instability}
\label{Mass-Transfer Instability}

As shown in Figure~\ref{fig:3-2-stable}, under the quasi-adiabatic criterion adopted in this work, stable mass transfer can still be maintained even when the mass-transfer rate reaches $\log \dot{M}_{\mathrm{d}} \sim -2$. This result is crucial for our study. In contrast, many previous studies have imposed relatively low upper limits on the mass-transfer rate for stability; for example, \citet{chenFormationCVntypeBinaries2017} and \citet{zhangImpactMassTransfer2021} adopted $\log \dot{M}_{\mathrm{d}} = -4$ as the maximum rate for stable mass transfer.

\begin{figure*}[htbp]
	\centering
	\begin{subfigure}[b]{0.49\textwidth}
		\centering
		\includegraphics[width=\textwidth]{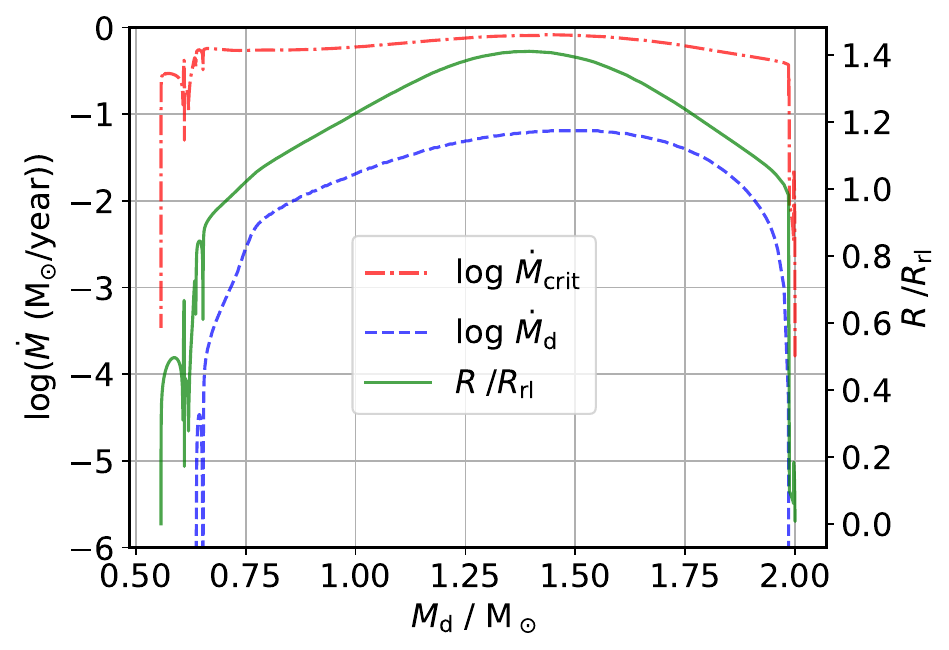}
		\caption{}
		\label{fig:3-2-outflow}
	\end{subfigure}
	\hfill
	\begin{subfigure}[b]{0.49\textwidth}
		\centering
		\includegraphics[width=\textwidth]{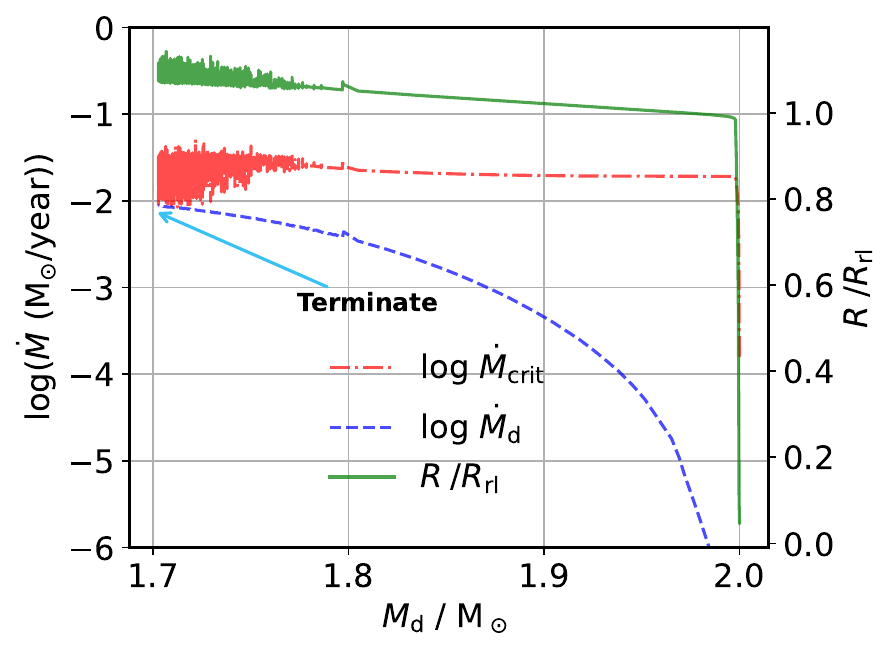}
		\caption{}
		\label{fig:3-2-unstable}
	\end{subfigure}
	
	\caption{These two panels show examples of the models that we exclude. They have the same initial donor mass ($2.0\,M_\odot$), accretor mass ($1.4\,M_\odot$), and initial metallicity ($10^{-1}\, Z_\odot$), but different initial orbital periods. Panel (a) has an initial orbital period of 450 days and represents a case in which the donor radius exceeds one-third of its Roche-lobe radius. Panel (b) has an initial orbital period of 30 days and illustrates a case of unstable mass transfer according to the quasi-adiabatic criterion.}
	\label{fig:3-2-exclude}
\end{figure*}

 \begin{figure*}[htbp]
	\centering
	\begin{subfigure}[b]{0.45\textwidth}
		\centering
		\includegraphics[width=\textwidth]{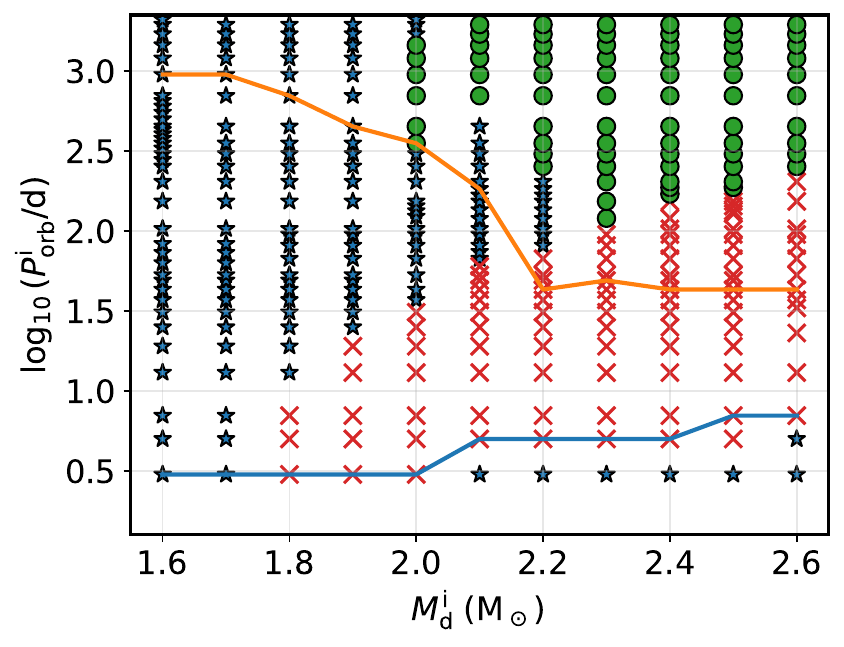}
		\caption{$Z^{\mathrm{i}} = Z_\odot$}
		\label{fig:3-3-a}
	\end{subfigure}
	\hfill
	\begin{subfigure}[b]{0.45\textwidth}
		\centering
		\includegraphics[width=\textwidth]{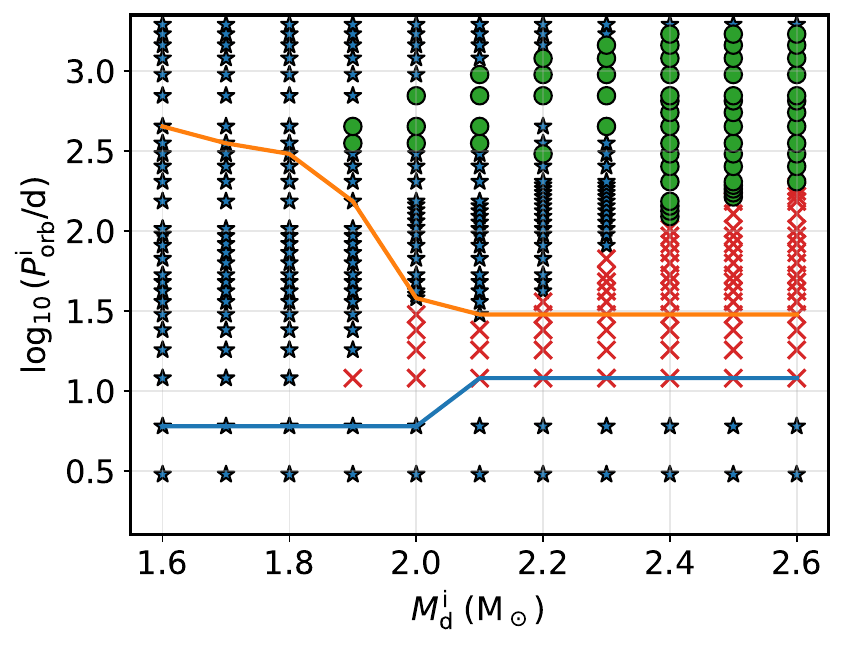}
		\caption{ $Z^{\mathrm{i}} = 10^{-1}\, Z_\odot$} 
		\label{fig:3-3-b}
	\end{subfigure}

	\begin{subfigure}[b]{0.45\textwidth}
		\centering
		\includegraphics[width=\textwidth]{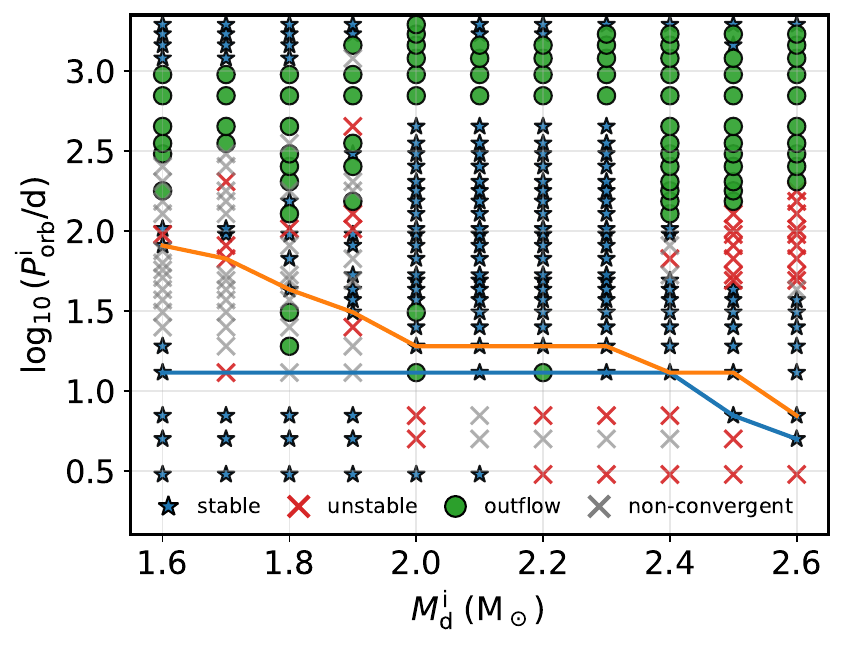}
		\caption{$Z^{\mathrm{i}} =10^{-3}\, Z_\odot$}
		\label{fig:3-3-c}
	\end{subfigure}
	\hfill
	\begin{subfigure}[b]{0.45\textwidth}
		\centering
		\includegraphics[width=\textwidth]{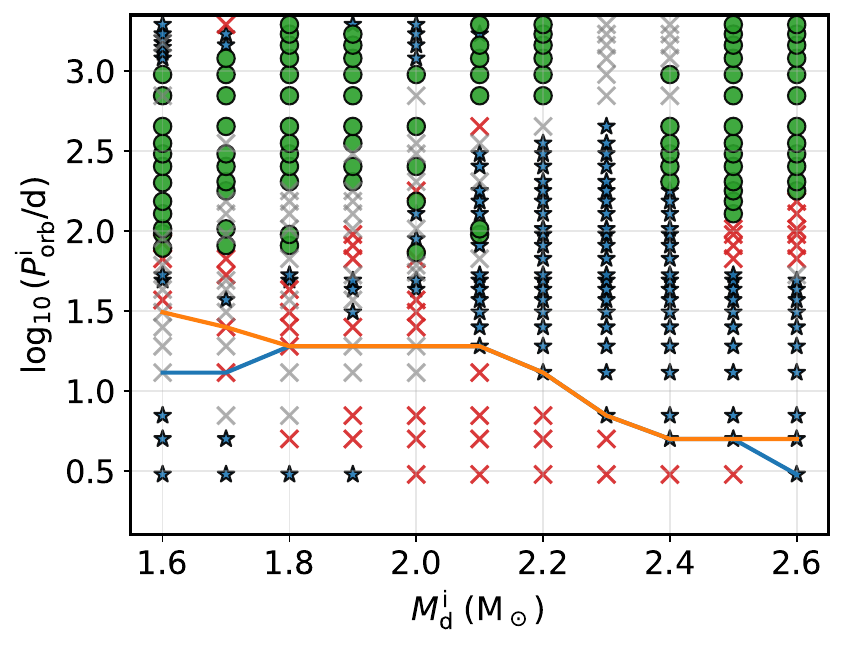}
		\caption{ $Z^{\mathrm{i}} = 10^{-4}\, Z_\odot$} 
		\label{fig:3-3-d}
	\end{subfigure}

	\caption{Comparison of the parameter space for stable mass transfer, unstable mass transfer, and outflow (mass loss through the outer Lagrange point $\mathrm{L}_3$) under different metallicities. All the computed models shown here have an accretor mass of $1.4\, M_{\odot}$, with periods ranging from 3 to 1950 days. The upper line shows the models with the minimum orbital periods leading to the onset of mass transfer in Case C, while the lower line shows the models with the minimum orbital periods for which mass transfer occurs after the main sequence and the Hertzsprung gap.}
	\label{fig:3-3-ab}
\end{figure*}

As shown in Figure~\ref{fig:3-2-exclude}, we present two models excluded from our analysis. 

In panel (a), the model is classified as stable under the quasi-adiabatic criterion. However, the donor star expands to $R\,/R_{\mathrm{rl}} \gtrsim 1.33$, which in our calculations typically corresponds to the donor radius exceeding the equivalent radius associated with the donor-side outer Lagrangian point.  We exclude this kind of model from our analysis because mass loss through the outer Lagrangian point may affect the final orbital periods, while such a mass-loss channel is not treated in our current MESA implementation in a self-consistent way.

Similar situations have been reported in previous studies. It is known from \citet{geAdiabaticMassLoss2020a} that some binary systems may undergo dynamically stable mass transfer on dynamical timescales. However, \citet{geThermalEquilibriumMassloss2020} showed that such systems can subsequently enter a thermal-timescale mass-transfer phase associated with outer-Lagrangian mass loss.

Panel (b) illustrates an instance of unstable mass transfer. In our calculations, mass transfer is considered unstable when the mass-transfer rate exceeds the local thermal-timescale mass-transfer rate defined in Equation~\ref{eq:crititcal rates}, indicating that the system cannot maintain stable mass transfer.

 Our results show that the initial metallicity has a significant impact on the parameter space in which stable mass transfer can occur in binary systems (Figure~\ref{fig:3-3-ab}).

For systems with a $1.4\,M_\odot$ accretor, the behavior of the models depends strongly on metallicity. For solar and one-tenth solar metallicities, donors with initial masses larger than $2.2\,M_\odot$ or $2.3\,M_\odot$ that initiate mass transfer in Case C, where the models located above the orange line in Figure~\ref{fig:3-3-ab}, generally evolve into instability, either violating the quasi-adiabatic criterion or leading to mass loss through the outer Lagrangian point $\mathrm{L}_3$. In contrast, for extremely low-metallicity models, stable mass transfer in Case C can still occur even for donors with initial masses up to $2.6\,M_\odot$.

As shown in Figure~\ref{fig:3-3-ab}, the orange curves in panels (c) and (d) lie significantly below those in panels (a) and (b), indicating that Case~C mass transfer occurs at shorter initial orbital periods and thus smaller stellar radii at lower metallicities.
For the same accretor mass and initial orbital period, these extremely low-metallicity models tend to have lower mass-transfer rates, which makes stable mass transfer more likely to occur.

This difference is the primary reason why the initial parameter space required for stable mass transfer in systems with intermediate-mass progenitors differs under extremely low-metallicity conditions.
We note that, for low-mass progenitors at extremely low metallicity, a large fraction of the models in our calculations failed to converge, preventing a reliable comparison with the higher-metallicity cases.

We have made a comparison between the Kolb mass-transfer scheme \citep{kolbComparativeStudyEvolution1990} and the Han mass-transfer scheme \citep{hanLowIntermediatemassClose2002,chenOrbitalPeriodsSubdwarf2013}, the latter being implemented within MESA using the \texttt{use\_other\_rlo\_mdot} option to enable an alternative mass-transfer prescription.
Compared with the Kolb scheme, the Han scheme yields systematically higher mass-transfer rates and is therefore more prone to entering unstable mass-transfer processes. In particular, for intermediate-mass donors, the Han scheme leaves almost no parameter space in which mass transfer can remain stable under the quasi-adiabatic criterion. Therefore, our study primarily focuses on the Kolb scheme.

In summary, both the metallicity and the adopted mass-transfer scheme affect the parameter space for stable mass transfer in binary systems, and consequently influence the mass–period parameter space under stable mass transfer.

\subsection{Stable $M_{\mathrm{WD}}$–$P_{\mathrm{orb}}$ Relation for low-mass donors}
\label{section 3.1}

In the Han scheme,
the mass-transfer rate depends only on the ratio between the donor
radius and its Roche-lobe radius ($R\,/R_{\mathrm{rl}}$),
and mass transfer starts only when this ratio exceeds unity
($R\,/R_{\mathrm{rl}} > 1$), which means that when the donor radius exceeds its Roche-lobe radius, mass is lost on a very short timescale. In contrast, the Kolb scheme accounts for several physical properties of the donor star, including the effective temperature, the pressure scale height at the photosphere, and the binary mass ratio, which together determine the mass-transfer rate through the donor's stellar atmosphere, allowing mass transfer
to occur even when the donor radius is slightly smaller than the
Roche-lobe radius.

\begin{figure*}[t]
\centering 
\includegraphics[width=0.9\textwidth]{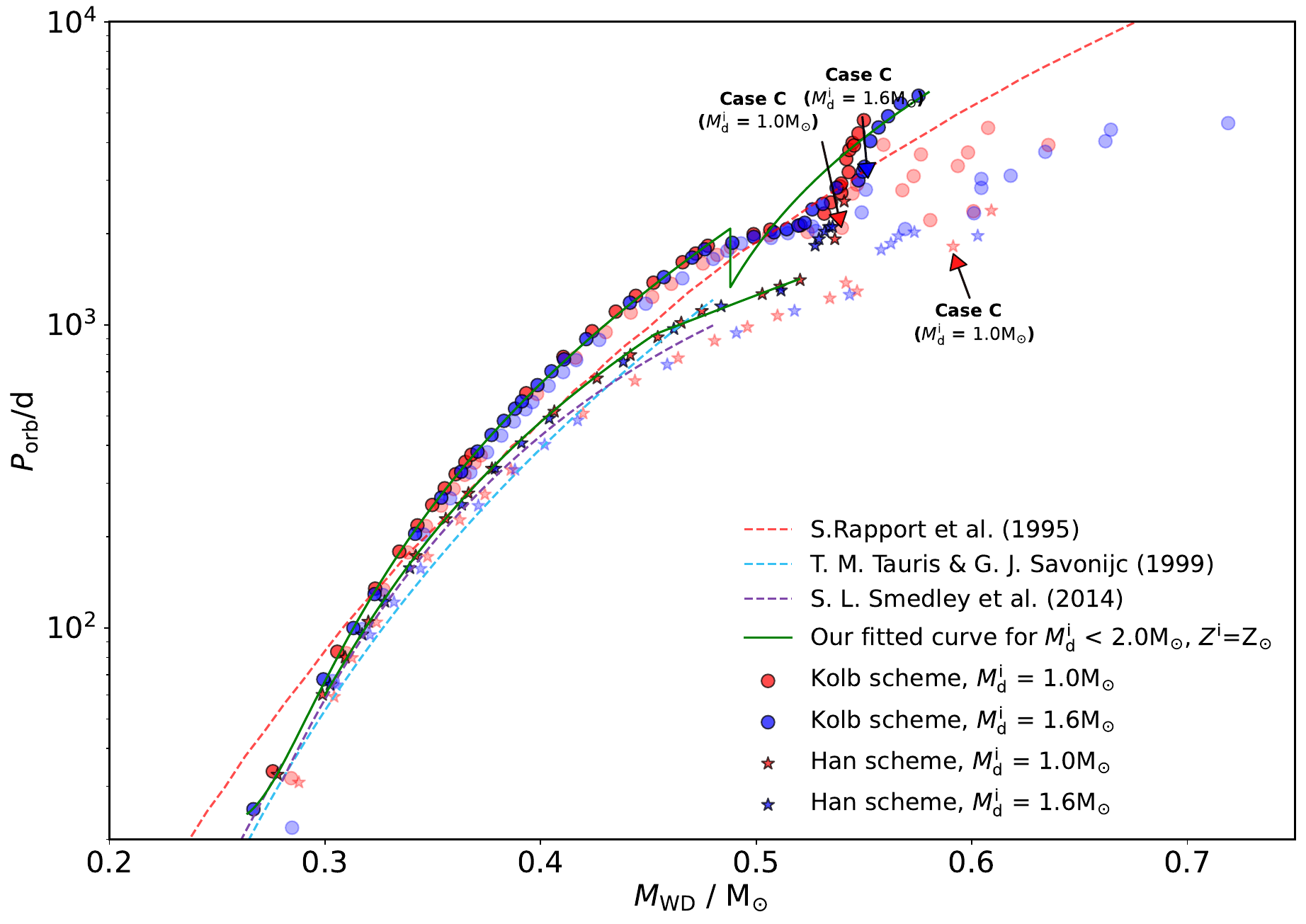} 
\caption{Comparison of the $M_{\text{WD}}$–$P_{\text{orb}}$ relationship between the Kolb scheme \citep{kolbComparativeStudyEvolution1990} and the Han scheme \citep{hanLowIntermediatemassClose2002,chenOrbitalPeriodsSubdwarf2013} for binary models with an initial  metallicity $Z^{\mathrm{i}} = Z_\odot$.
Light-colored symbols indicate models at the end of the mass-transfer phase, whereas dark-colored symbols represent models during the white dwarf cooling phase.
The solid circles and asterisks represent results obtained using the Kolb and Han mass-transfer schemes, respectively. Red and blue symbols denote models with initial donor masses of $1.0 \, M_\odot$ and $1.6 \, M_\odot$, respectively. The green lines show the analytical fits to these numerical models based on Equation \ref{eq}.}
\label{fig-kvc} 
\end{figure*}

\begin{table*}[htbp]
\centering
\caption{Fitted parameters for Equation~\ref{eq} for the case of stable $M_{\mathrm{WD}}$--$P_{\mathrm{orb}}$ relation.}
\label{tab:2}
\begin{tabular}{cccccc}
\hline
$M_{\mathrm{WD}}/M_\odot$ &Metallicity& Mass-transfer scheme & $a$ & $b$ & $c$ \\
\hline
0.264-0.488& $Z_\odot$ & Kolb & $1.586$ & $5.356$ & $-7.150$ \\

0.488-0.600& $Z_\odot$ & Kolb & $0.608$ & $4.671$ & $-39.274$ \\

0.303-0.530& $10^{-1} \, Z_\odot$ & Kolb & $0.943$ & $3.373$ & $-9.737$ \\

0.530-0.670& $10^{-1} \, Z_\odot$ & Kolb & $0.319$ & $4.952$ & $-105.458$ \\

0.338-0.392& $10^{-3} \, Z_\odot$ & Kolb & $1.408$ & $3.389$ & $-6.085$ \\

0.392-0.0.518& $10^{-3} \, Z_\odot$ & Kolb & $1.556$ & $3.069$ & $-5.035$ \\

0.364-0.440& $10^{-4} \, Z_\odot$ & Kolb & $0.888$ & $2.819$ & $-8.599$ \\

0.440-0.520& $10^{-4} \, Z_\odot$ & Kolb & $1.895$ & $2.973$ & $-4.066$ \\

0.300-0.453& $ \, Z_\odot$ & Han & $0.763$ & $3.628$ & $-13.989$ \\

0.453-0.520& $ \, Z_\odot$ & Han & $0.605$ & $3.279$ & $-14.786$ \\
\hline

\end{tabular}
\end{table*}

The two schemes will lead to a distinct $M_{ \mathrm{WD}}$--$P_{ \mathrm{orb}}$ relation (Figure~\ref{fig-kvc}). 
The line that represents the $M_{ \mathrm{WD}}$--$P_{ \mathrm{orb}}$ relation from the Han scheme is below the line that represents the relation from \citet{rappaportRelationWhiteDwarf1995}, which is consistent with the result of \citet{chenOrbitalPeriodsSubdwarf2013}. 

This regime corresponds to atmospheric mass transfer, in which matter flows through the stellar atmosphere before the donor fully fills its Roche lobe \citep{ritterTurningMassTransfer1988}.
That implies that the donor to which we apply the Han scheme has more time to increase its core mass before it starts to transfer mass. This is the main reason for the difference between the $M_{ \mathrm{WD}}$--$P_{ \mathrm{orb}}$ relations in the Kolb and the Han schemes.

We propose a formula that provides a good fit to the $M_{ \mathrm{WD}}-P_{ \mathrm{orb}}$ relation:
\begin{equation}
	P_{\text{orb}}=10^3 (e^{b}M^{a}_{\text{WD}}+cM_{\text{WD}}),
\label{eq}
\end{equation}
 where $a$, $b$, and $c$ are free parameters, as listed in Table~\ref{tab:2}, and $P_{\text{orb}}$ and $M_{\text{WD} }$ denote the periods (in days) and donor masses (in solar masses) at the WD cooling stages. The donor masses at the end of their mass-transfer phases sometimes do not represent the final masses of their WD cooling stages, because their mass will decrease owing to stellar winds during their later evolution. 
This discrepancy is generally larger in long-period binaries (see Figure~\ref{fig-kvc}). At the end of mass transfer, the donor stars in such systems detach from their Roche lobes at relatively large radii, implying more massive residual envelopes and consequently stronger wind-driven mass loss during subsequent evolution.

\begin{figure*}[htbp]
	\centering
	\begin{subfigure}[b]{0.49\textwidth}
		\centering
		\includegraphics[width=\textwidth]{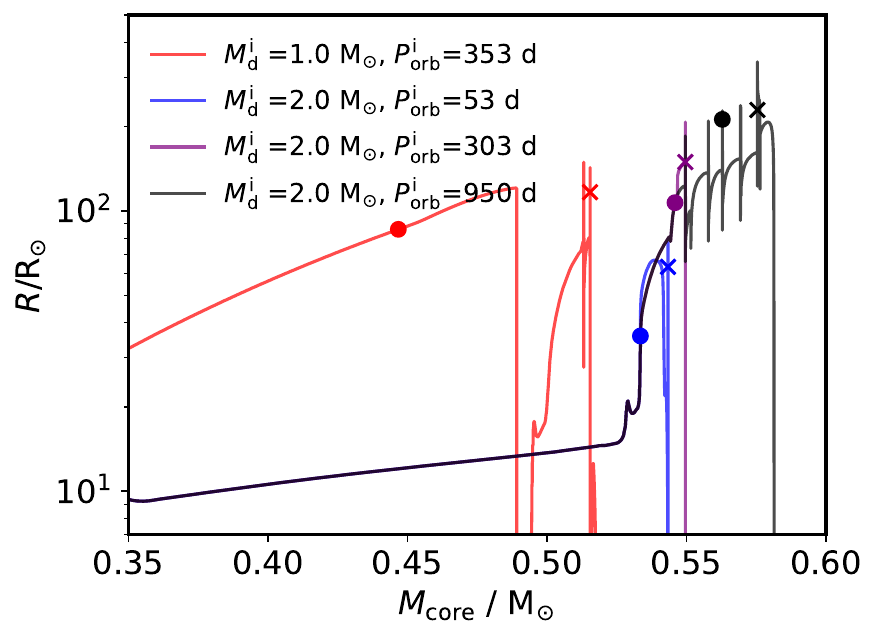}
		\caption{}
		\label{fig:3-2-b}
	\end{subfigure}
	\hfill
	\begin{subfigure}[b]{0.49\textwidth}
		\centering
		\includegraphics[width=\textwidth]{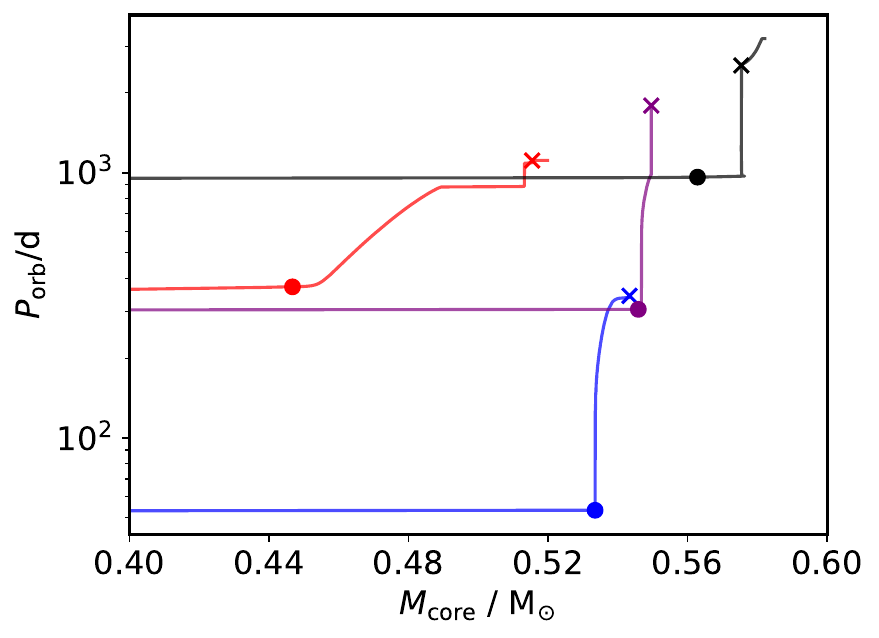}
		\caption{}
		\label{fig:3-2-c}
	\end{subfigure}
	
\caption{ 
Evolution of binary models with one-tenth solar metallicity and an accretor mass of $1.4\,M_\odot$. The left panel shows the donor radius as a function of core mass, while the right panel shows the orbital period as a function of core mass. Filled circles mark the onset of mass transfer and crosses indicate the end of the mass-transfer phase, defined by $\log \dot{M}_{\mathrm d}=-9$. Both the donor radius and orbital period are plotted on logarithmic scales.
}
	\label{fig:3-2-bc}
\end{figure*}

\begin{figure}[htbp] 
	\centering \includegraphics[width=\columnwidth]{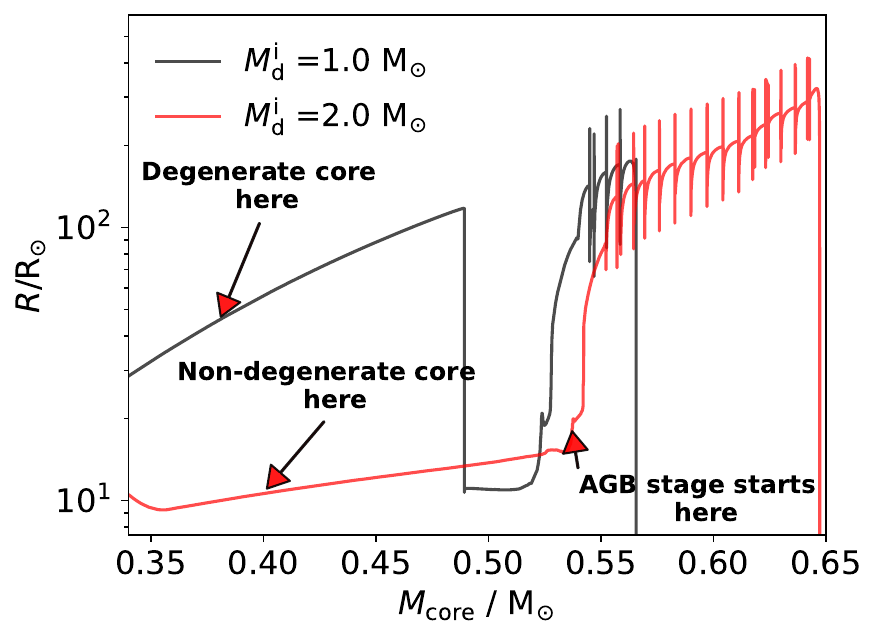} 
	\caption{Relation between the core mass and stellar radius during single-star evolution for two models with initial masses of $1.0\,M_\odot$ and $2.0\,M_\odot$ at one-tenth solar metallicity.}
	\label{fig:3-2-sgs} 
\end{figure}

Various metallicities can produce different $M_{\mathrm{WD}}$--$P_{\mathrm{orb}}$ relations \citep{rappaportRelationWhiteDwarf1995,chenOrbitalPeriodsSubdwarf2013}.
For extremely low-metallicity models, when the initial mass is around $1.7\,M_\odot$, the core mass–radius relation is no longer obeyed.
We can also fit these $M_{\mathrm{WD}}$--$P_{\mathrm{orb}}$ relations for the models with the initial metallicity of $10^{-3}\, Z_\odot$ and $10^{-4}\, Z_\odot$ originating form progenitors with degenerate cores by Equation (\ref{eq}). The relevant fitted parameters are listed in Table~\ref{tab:2}. 


\subsection{$M_{\mathrm{WD}}$–$P_{\mathrm{orb}}$ Distribution for intermediate-mass donors}
\label{sec-Distribution for Progenitors with intermediate mass}

The properties of the cores of WD progenitors significantly affect the 
$M_{\mathrm{WD}}$--$P_{\mathrm{orb}}$ relation. Low-mass progenitors 
typically develop degenerate helium cores, leading to a well-defined 
core mass-radius relation. In contrast, intermediate-mass stars ignite 
helium under nondegenerate conditions. However, after central helium 
burning, they form degenerate C/O cores \citep{pols2011stellar}.

As illustrated in Figure~\ref{fig:3-2-b}, the evolutionary behavior of the donor's radius is intrinsically linked to the degeneracy of its core. For the low-mass model (red curve), the helium core is highly degenerate, and the stellar radius follows a strict core mass--radius relation where $R$ increases monotonically with $M_{\mathrm{core}}$. In contrast, for intermediate-mass models where the core remains nondegenerate until the formation of a C/O core, the stellar radius shows only weak evolution over a significant portion of the core helium-burning phase, despite substantial core growth.
Consequently, in intermediate-mass donors, a much larger fraction of the final core mass is accumulated prior to the onset of mass transfer. This difference in radial expansion has a profound impact on the binary's initial configuration. To reach a specific core mass at the onset of mass transfer, a donor with a nondegenerate core requires a significantly shorter initial orbital period than a low-mass donor with a degenerate core. Since the dramatic increase in the orbital period primarily occurs during the mass-transfer phase (as shown in Figure~\ref{fig:3-2-c}), the final period is heavily influenced by this initial state. Specifically, Figure~\ref{fig:3-2-c} compares a $2\,M_\odot$ donor (initial $P_{\mathrm{orb}} = 53$ days, blue curve) with a $1.0\,M_\odot$ donor (initial $P_{\mathrm{orb}} = 353$ days, red curve). While both systems eventually reach the comparable final core masses, the intermediate-mass donor's substantially shorter initial period prevents it from reaching the long orbital periods characteristic of the low-mass $M_{\mathrm{WD}}$--$P_{\mathrm{orb}}$ relation. As a result, its final position in the $M_{\mathrm{WD}}$--$P_{\mathrm{orb}}$ plane lies notably below the canonical relation.

As illustrated in Figure~\ref{fig:3-2-sgs}, during the thermal pulsing phase, a single star of $2.0\, M_{\odot}$ exhibits a core mass–radius relation similar to that of a $1.0\, M_{\odot}$ single star.
The significant core growth for intermediate-mass stars during the AGB phase occurs after several thermal pulses have taken place. As shown by the red curve in Figure~\ref{fig:3-2-sgs}, during the evolution of a 2.0 $ M_{\odot}$ single star, while its radius increases from 16~$\rm R_{\odot}$ to 195 $ R_{\odot}$ (the first radius peak), the core mass only grows by approximately 0.02 $M_{\odot}$.
As the thermal pulses proceed, the stellar radius expands to progressively larger values. Therefore, a substantially larger orbital separation would be required for Roche-lobe overflow to occur after many thermal pulses.

As the orbital period increases, the intermediate-mass donors become able to start mass transfer after the first peak of the thermal pulsing phase and their cores are more degenerate, which follows the core mass–radius relation characteristic of low-mass progenitors. As a result, the corresponding systems occupy a region of the $M_{\mathrm{WD}}$--$P_{\mathrm{orb}}$ plane that approaches the relation produced by low-mass donors.

\begin{figure*}[htbp]
	\centering 
	\includegraphics[width=0.9\textwidth]{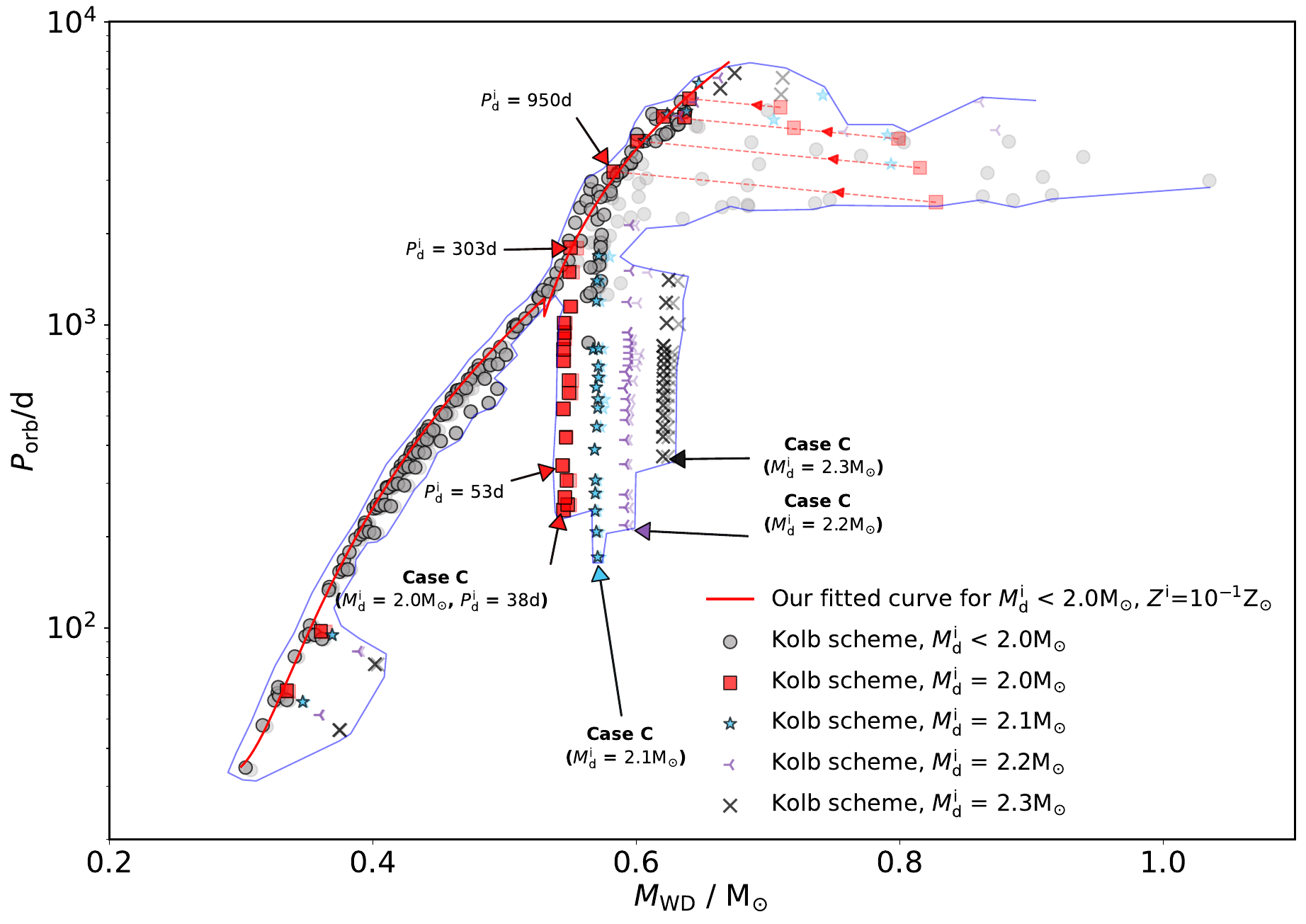} 
	\caption{
Comparison of the $M_{\mathrm{WD}}$--$P_{\mathrm{orb}}$ distributions for low-mass and intermediate-mass progenitors with an initial metallicity of $10^{-1}\,Z_\odot$. 
The color shade distinguishes different evolutionary phases: lighter markers correspond to the end of mass transfer, and darker markers correspond to the subsequent WD cooling phase. Pairs of points connected by dashed arrows correspond to the same models at different evolutionary stages.
The red curve shows the fitted relation (Equation ~\ref{eq}), which represents the $M_{\mathrm{WD}}$--$P_{\mathrm{orb}}$ relation for low-mass progenitors. 
The models span an initial orbital period range from 3 to 1950 days; however, only systems undergoing stable mass transfer are shown. Systems that experience unstable mass transfer or mass loss through the outer Lagrange point $\mathrm{L}_3$ are excluded.
}
	\label{fig-3-2-a} 
\end{figure*}

Take the models with initial mass of $2.0\,M_\odot$ and initial metallicity of $10^{-1}\, Z_\odot$ as an example (the red points in Ficture ~\ref{fig-3-2-a}). 
When the initial orbital period ranges from 38 to 303 days, variations in the initial orbital period result in only negligible changes in the core mass prior to the onset of mass transfer. 
When the initial orbital period is shorter than 303 days ($P_{\mathrm{orb}}^{\mathrm{i}}  \leq 303 \, \rm days$), increasing the orbital period does not allow the donors to initiate mass transfer after the first radius peak associated with thermal pulsing. 
This implies that the donors accumulate comparable core masses at the onset of mass transfer.
At an initial orbital period of 950 days, the donor can initiate mass transfer after its radius reaches the first peak during the thermal pulsing phase (see the black filled circle in Figure~\ref{fig:3-2-b}). 
Therefore, for initial orbital periods longer than 950 days ($P_{\mathrm{orb}}^{\mathrm{i}}  \geq 950 \, \rm days$), the $M_{\mathrm{WD}}$--$P_{\mathrm{orb}}$ distribution is no longer concentrated around $0.55\,M_\odot$.

When the initial orbital period periods range from $38$ to $303$ days 
($38 \, \mathrm{days} \le P_{\mathrm{orb}}^{\mathrm{i}}
\le 303 \, \mathrm{days}$), variations in the initial orbital period mainly affect the final orbital period, while having little impact on the mass of the resulting WDs. As a consequence, the $M_{\mathrm{WD}}$--$P_{\mathrm{orb}}$ distribution exhibits an approximately vertical trend.

\begin{figure*}[htbp]
	\centering 
	\includegraphics[width=0.9\textwidth]{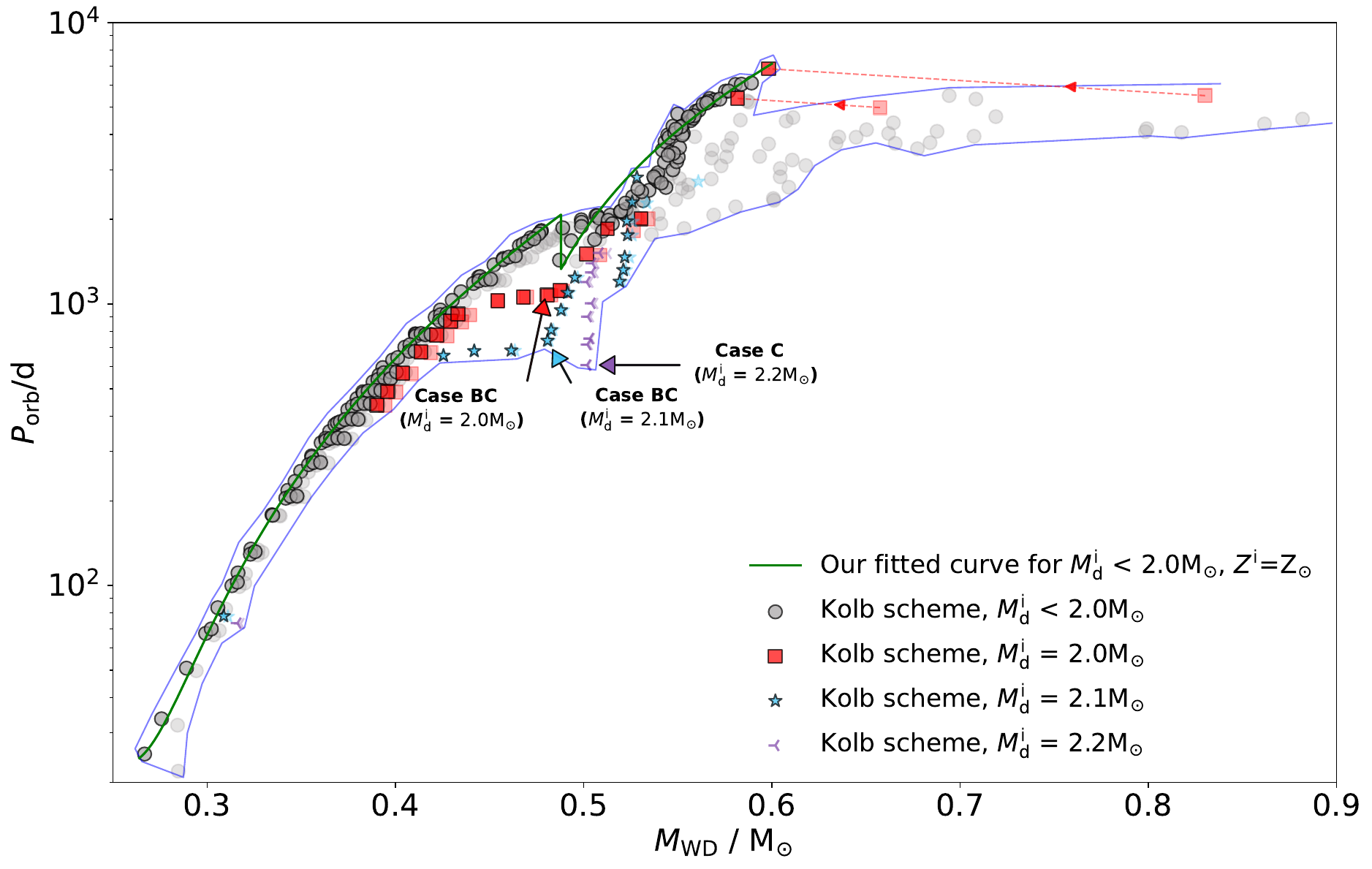} 
	\caption{
Same as Figure~\ref{fig-3-2-a}, but for an initial metallicity of $Z_\odot$. The models span an initial orbital period range from 3 to 1950 days; only systems undergoing stable mass transfer are shown. Systems with unstable mass transfer or mass loss through the outer Lagrange point $\mathrm{L}_3$ are excluded. The red curve shows the fitted relation (Equation ~\ref{eq}), representing the $M_{\mathrm{WD}}$--$P_{\mathrm{orb}}$ relation for low-mass progenitors.}
	\label{fig-3-2-a0} 
\end{figure*}

Overall, the $M_{\mathrm{WD}}$--$P_{\mathrm{orb}}$ distribution produced by intermediate-mass donors lies below that produced by low-mass donors in both the $0.1\,Z_\odot$ and solar-metallicity cases (Figures~\ref{fig-3-2-a} and \ref{fig-3-2-a0}).

In the solar-metallicity models shown in Figure~\ref{fig-3-2-a0}, a significant deviation from the standard relation begins to appear when the initial donor mass reaches $2.1\,M_\odot$, whereas the $2.0\,M_\odot$ models still remain close to the canonical low-mass progenitor relation.

According to \citet{marigoZerometallicityStarsEvolution2001} and \citet{karakasDawesReview22014}, 
for stars with solar metallicity, the core degeneracy during the RGB phase becomes insufficient to support the occurrence of a helium flash when the stellar mass reaches $2.1\,M_\odot$, while for extremely metal-poor stars with $\mathrm{[Fe/H]}$ = $-2.3$, the corresponding transition occurs at a lower mass of approximately $1.75\,M_\odot$.
This is consistent with our results. 
Influenced by the evolution of core degeneracy in the donor star, our models with solar metallicity show a significant deviation from the standard $M_{\mathrm{WD}}$--$P_{\mathrm{orb}}$ relation for stable mass transfer when the initial mass is $2.1\,M_\odot$ and above.

Figures~\ref{fig-3-2-a} and \ref{fig-3-2-a0} show only those systems that undergo stable mass transfer within the explored initial orbital-period range of 3--1950\,days. 
For the intermediate-mass donor models shown in Figures~\ref{fig-3-2-a} and \ref{fig-3-2-a0} (the arrows indicate example Case C/BC models, illustrating that, for a given donor mass, systems with longer final orbital periods undergo stable mass transfer during the AGB phase), we include systems undergoing stable mass transfer during the Hertzsprung gap, the RGB, and the AGB phases. In particular, among our calculated models, systems with $M_{\mathrm{WD}}$ around $0.31 \,M_\odot$ at solar metallicity and around $0.38 \,M_\odot$ at one-tenth solar metallicity undergo mass transfer that begins in the Hertzsprung gap and continues during the RGB phase. Models undergoing stable mass transfer before the AGB phase likewise deviate from the canonical low-mass-donor $M_{\mathrm{WD}}$--$P_{\mathrm{orb}}$ relation because their donor stars follow different core mass--radius relations at these evolutionary stages.

We note that the range of donor masses shown ($2.0 \text{--} 2.3 \, M_{\odot}$) is limited by the mass-transfer instability. This is primarily due to the increasing initial mass ratio  $q$, since the companion mass is kept fixed in these models.

Specifically, for models with solar metallicity, systems with donor masses $\text{M}_{\mathrm{d}}^{\mathrm{i}} > 2.2 \, M_{\odot}$ typically undergo unstable mass transfer or experience significant mass loss through the outer Lagrange points $L_3$. However, for the lower-metallicity cases ($0.1 \, Z_{\odot}$), this stability threshold extends slightly higher to $\text{M}_{\mathrm{d}}^{\mathrm{i}} > 2.3 \, M_{\odot}$, owing to the more compact structures of metal-poor stars. Systems exceeding these respective mass limits do not produce the stable remnants shown in our $M_{\mathrm{WD}} \text{--} P_{\mathrm{orb}}$ parameter space.

A notable feature in our results is the metallicity dependence of the final orbital periods. At lower metallicity ($Z^{\mathrm{i}} = 0.1 \, Z_{\odot}$), systems can initiate stable mass transfer at shorter periods during the AGB phase compared to solar-metallicity models. This is primarily because low-metallicity stars remain more compact—having smaller stellar radii—than their metal-rich counterparts at the same evolutionary stage. Consequently, these systems can satisfy the Roche-lobe overflow condition at closer separations, leading to the smaller final periods observed in the $M_{\mathrm{WD}}$--$P_{\mathrm{orb}}$ plane.

\begin{figure*}[htbp]
	\centering 
	\includegraphics[width=0.9\textwidth]{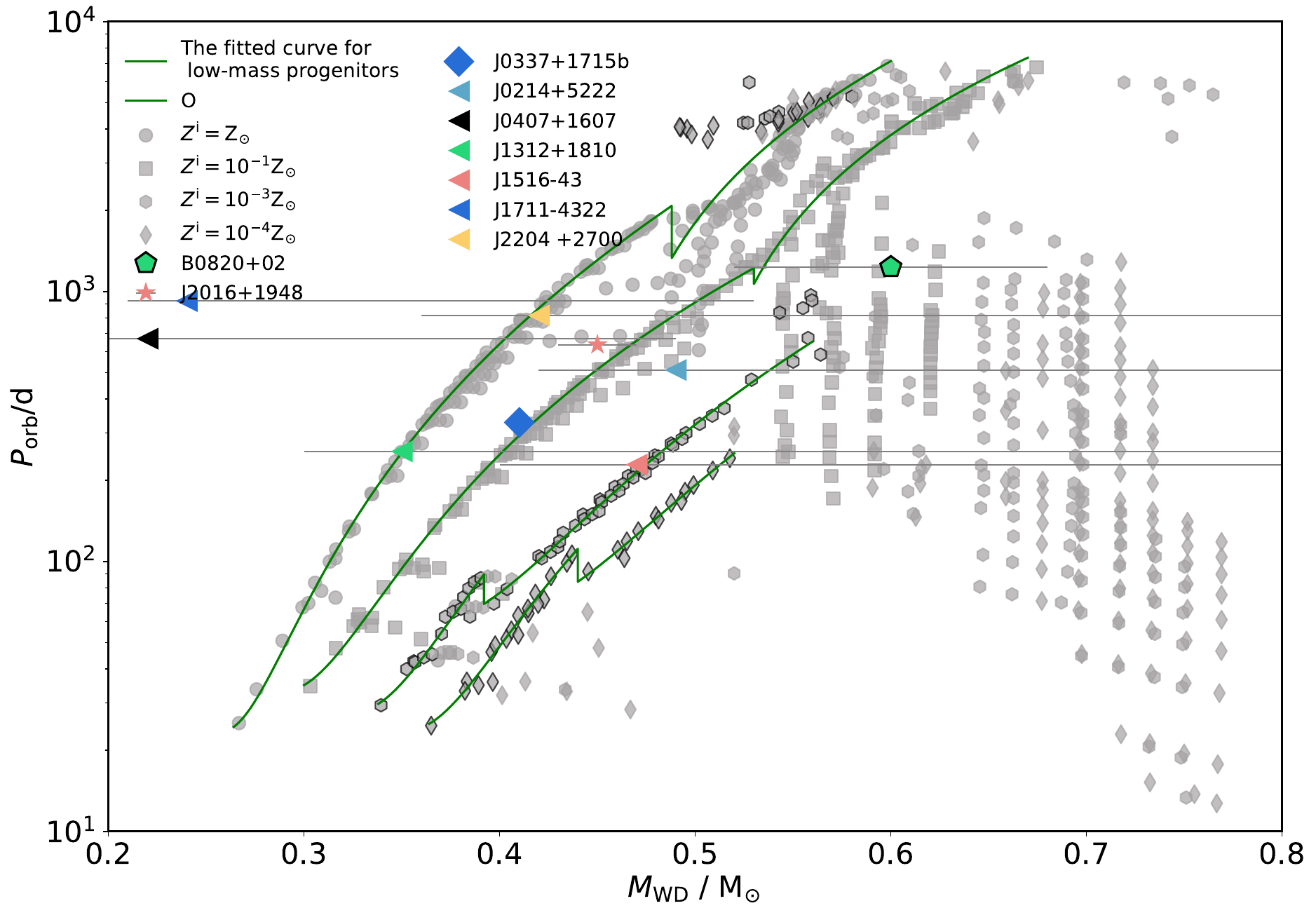} 
	\caption{Comparison between observations and theoretical models. The accretor masses are $1.4 \, M_\odot$. 
		Computed models outlined with borders denote the extremely metal-poor cases that exhibit a clear $M_{\mathrm{WD}}$--$P_{\mathrm{orb}}$ relation. These correspond to models with $M_{\mathrm{d}}^{\mathrm{i}} < 1.7\,M_\odot$ for $10^{-3}\,Z_\odot$, and $M_{\mathrm{d}}^{\mathrm{i}} < 1.6\,M_\odot$ for $10^{-4}\,Z_\odot$.
		All observed systems shown are  WD+NS binaries. Observed sources marked with triangles are from  \citet{manchesterAustraliaTelescopeNational2005}. The pentagon with a black border represents the C/O WD binary B0820+02, while all other observed markers denote He WD binaries. Gray points in the background represent our calculated models. The green lines represent the fitted $M_{\mathrm{WD}}$--$P_{\mathrm{orb}}$ relations for low-mass progenitors. From top to bottom, the lines correspond to metallicities of $Z_\odot$, $10^{-1}\, Z_\odot$, $10^{-3}\, Z_\odot$, and $10^{-4}\, Z_\odot$, respectively.}
	\label{fig-3-4-a} 
\end{figure*}

The physical origin of the deviations from the canonical $M_{\mathrm{WD}}$--$P_{\mathrm{orb}}$ relation shown by our models, as well as by the observed systems discussed in the following section, can be understood within the framework presented in this section.
The only exceptions are the long-period, extremely metal-poor models shown as black-bordered symbols in Figure~\ref{fig-3-4-a}. 
According to the canonical $M_{\mathrm{WD}}$--$P_{\mathrm{orb}}$ relation for low-mass donors, longer final orbital periods correspond to more massive WD. However, for extremely metal-poor low-mass donors, a phase of rapid radial expansion occurs at the onset of the AGB phase, during which the core mass changes only slightly. During this phase, the donor radius exceeds the maximum radius reached during the RGB phase, leading to Roche-lobe overflow at much larger orbital separations without a corresponding increase in the WD mass. Consequently, some low-mass-donor models in Figure~\ref{fig-3-4-a} deviate from the canonical $M_{\mathrm{WD}}$--$P_{\mathrm{orb}}$ relation, with many of them reaching final orbital periods exceeding $1000$ days.

It is worth noting that the apparent gaps in the orbital-period distributions, in particular the prominent gap between the short- and long-period extremely metal-poor models in Figure~\ref{fig-3-4-a}, are partly due to the exclusion of models that encountered numerical convergence problems, underwent unstable mass transfer, or overflowed beyond the outer Lagrangian surface.

\subsection{Comparison with Observation}
In this section, we compare our models with several observed WD binaries (the parameters of the individual observed systems are listed in Table~\ref{tab:1}), focusing on systems that deviate from the canonical $M_{\mathrm{WD}}$--$P_{\mathrm{orb}}$ relation. These systems provide important constraints on the formation channels explored in this work, in particular for long-period binaries and systems located below the standard relation.

Figures~\ref{fig-3-4-a} and ~\ref{fig-5-c} compare our theoretical models with the observed WD+NS and WD+MS binaries, respectively. We focus in particular on the observed systems that deviate from the canonical $M_{\mathrm{WD}}$--$P_{\mathrm{orb}}$ relation.

\subsubsection{WD+NS Binaries}

In our observational comparison, we include several WD+NS systems in which the neutron-star companions are observed as radio pulsars. The corresponding pulsar identifications can be found in \citet{manchesterDetectionPulsarLongperiod1980}, \citet{manchesterAustraliaTelescopeNational2005}, \citet{navarroArecibo430MHzIntermediate2003}, and \citet{ransomMillisecondPulsarStellar2014}.
Among them are six He WD-NS binary systems selected from \citet{manchesterAustraliaTelescopeNational2005}:
J0214+5222, J0407+1607, J1312+1810, J1516$--$43, J1711$--$4322, and J2204+2700 (marked by left-pointing triangles in Figure~\ref{fig-3-4-a}).
The WD parameters reported by \citet{manchesterAustraliaTelescopeNational2005} suffer from significant uncertainties in their upper mass limits, primarily due to the unknown orbital inclination angles. In that work, all pulsars in these systems are assumed to have a mass of $1.35 \, M_\odot$, from which the median WD's masses are derived. Among these systems, two WDs (J0407+1607 and J0214+5222) appear to lie above our model predictions, which is likely a consequence of the uncertainty in the orbital inclination angles.
Within the observational uncertainties, all six systems can be explained by the $M_{\mathrm{WD}}$--$P_{\mathrm{orb}}$ relations corresponding to low-mass progenitors.
Among these systems, J1516$-$43 is particularly noteworthy. 
Its observational uncertainty only marginally overlaps with the $M_{\mathrm{WD}}$--$P_{\mathrm{orb}}$ relation for low-mass progenitors at a metallicity of $10^{-1} \, Z_\odot$. 
Instead, J1516$-$43 is located within the parameter space corresponding to stable mass transfer from extremely metal-poor, low-mass progenitors.

J0337+1715 is a hierarchical triple system consisting of a neutron star and two He WD.
J0337+1715b lies between the mass–period relations predicted for low-mass progenitors with $ Z_\odot$ and $10^{-1} \, Z_\odot$ (see Figure~\ref{fig-3-4-a}).

\begin{table*}[htbp]
	\centering
	\caption{Observed WD Binary Samples}
	\label{tab:1}
	\begin{tabular}{cccccc}
		\hline
		name & type & period & $M_{\mathrm{1}}$ & $M_{\mathrm{2}}$ & references \\
		\hline
		 &  & (days) & $M_\odot$ & $M_\odot$ & \\
		\hline
		J1836--5110 & HeWD+subgiant & 461.48 & $0.40^{+0.01}_{-0.01}$ & $1.38^{+0.16}_{-0.16}$ & \citet{parsonsWhiteDwarfBinary2022} \\
        
		BD$-$22$^\circ$3467 & WD+subgiant & 790.22 & $0.53^{+0.040}_{-0.025}$ & $1.1^{+0.2}_{-0.2}$ & \citet{lobling2020first} \\

		B0820+02 & C/OWD+NS & 1232.5 & $0.6^{+0.08}_{-0.08}$ & - & \citet{koester2000white} \\
		J2016+1948 & HeWD+NS & 635.0238 & $0.45^{+0.02}_{-0.02}$ & $1.0^{+0.5}_{-0.5}$ & \citet{gonzalezHIGHPRECISIONTIMINGFIVE2011} \\
		J0337+1715b & HeWD+HeWD+NS \footnote{This is a triple-star system.} & 327.3  & $0.4101^{+0.0003}_{-0.0003}$ & $1.4378^{+0.0013}_{-0.0013}$ & \citet{ransomMillisecondPulsarStellar2014} \\
		
		J0214+5222 & HeWD+NS & 512.0 & $0.48^{+0.70}_{-0.07}$ & - & \citet{manchesterAustraliaTelescopeNational2005} \\
		
		J0407+1607 & HeWD+NS & 669.1 & $0.22^{+0.27}_{-0.003}$ & - &  \citet{manchesterAustraliaTelescopeNational2005} \\
		
		J1312+1810 & HeWD+NS\footnote{After this paper was accepted, we found the update mass by \citet{lian2025thirty}} & 255.8 & $0.35^{+0.47}_{-0.05}$ & - &  \citet{manchesterAustraliaTelescopeNational2005} \\
		
		J1516--43 & HeWD+NS & 228.5 & $0.47^{+0.68}_{-0.07}$ & - &  \citet{manchesterAustraliaTelescopeNational2005} \\
		
		J1711--4322 & HeWD+NS & 922.5 & $0.24^{+0.29}_{-0.03}$ & - &  \citet{manchesterAustraliaTelescopeNational2005} \\

		J2204+2700 & HeWD+NS & 815.2 & $0.42^{+0.58}_{-0.06}$ & - &  \citet{manchesterAustraliaTelescopeNational2005} \\

		KIC 03835482 & C/OWD+MS & 683.267 & $0.53^{+0.08}_{-0.03}$ & $1.2^{+0.3}_{-0.1}$ &  \citet{kawaharaDiscoveryThreeSelflensing2018} \\
		
		KIC 06233093 & C/OWD+MS & 727.98 & $0.62^{+0.04}_{-0.05}$ & $1.2^{+0.2}_{-0.1}$ &  \citet{kawaharaDiscoveryThreeSelflensing2018} \\
		
		KIC 12254688 & C/OWD+MS & 418.715 & $0.62^{+0.09}_{-0.06}$ & $1.5^{+0.1}_{-0.1}$ &  \citet{kawaharaDiscoveryThreeSelflensing2018} \\
		\hline

	\end{tabular}
\end{table*}

As discussed in Section~\ref*{section 3.1}, the $M_{\mathrm{WD}}$–$P_{\mathrm{orb}}$ relation produced by low-mass progenitors shifts downward as the metallicity decreases. Therefore, J0337+1715b may originate from a progenitor with a metallicity between $Z = Z_\odot$ and $Z = 10^{-1} Z_\odot$, although further modeling would be required in order to test this possibility.
A similar situation also applies to J2016+1948.

B0820+02 is a binary system consisting of a C/O WD and an NS. 
Notably, only a small fraction of its observational uncertainty
overlaps with the stable $M_{\mathrm{WD}}$--$P_{\mathrm{orb}}$ relation. 
According to our model results, B0820+02 is more likely to have evolved from an intermediate-mass progenitor.

\subsubsection{WD+MS Binaries}
\label{susubsection obser of WD+MS}

We select several long-period binary systems with orbital periods longer than $200 \, \mathrm{days}$ for comparison with our models. Among them are three WD+MS binary systems from \citet{kawaharaDiscoveryThreeSelflensing2018}, all of which have MS companions with masses of approximately $1.4 \, M_\odot$. These systems are self-lensing binaries, and their WD companions all possess C/O cores. The stellar parameters adopted by \citet{kawaharaDiscoveryThreeSelflensing2018}, based on the KIC DR25 catalog \citep{mathurRevisedStellarProperties2017}, do not indicate extremely metal-poor compositions for these systems.

We also consider BD$-$22$^\circ$3467 \citep{mccook1999catalog}, a WD+subgiant binary with a mass and orbital period distribution similar to those of the self-lensing binaries. According to \citet{bhattacharjee2026thermally}, this WD does not exhibit extremely metal-poor composition. Its mass is adopted from \citet{lobling2020first} and its orbital period from \citet{bhattacharjee2026thermally}. The WD in this system has an uncertain core composition; it could be either a helium core or a carbon–oxygen-core.

In addition, we also consider J1836$-$5110, a binary system consisting of a He WD and a subgiant companion, since it provides an important observational example for testing the stable $M_{\mathrm{WD}}$--$P_{\mathrm{orb}}$ relation for low-mass progenitors at long orbital periods.

To compare our results with those for MS accretors, we computed intermediate-mass models with $\beta = 0.5$. The comparison is shown in Figure~\ref{fig-5-c}. In these calculations, neither the Eddington mass transfer limit nor the effects of X-ray irradiation on the mass-transfer efficiency are taken into account.
Notably, for an accretor mass of $1.4\,M_\odot$, the solar-metallicity models fail to reproduce the $M_{\mathrm{WD}}$--$P_{\mathrm{orb}}$ parameter space of the observed long-period WD+MS binaries shown in Figure~\ref{fig-5-c}. This is because, at solar metallicity, the range of initial orbital periods that allows stable mass transfer for intermediate-mass donors is relatively narrow, and the resulting final orbital periods cannot extend to the lower-period regime, as discussed in Section~\ref{sec-Distribution for Progenitors with intermediate mass}.
Our calculations further indicate that, at one-tenth solar metallicity, nearly all models with donor masses of $2.2\,M_\odot$ and above experience either unstable mass transfer or mass loss through the outer Lagrangian points.

Among the five observed systems considered here, only KIC 03835482 and J1836$--$5110 are consistent with the canonical WD $M_{\mathrm{WD}}$--$P_{\mathrm{orb}}$ relation for low-mass progenitors.

The remaining three systems, KIC 06233093, KIC 12254688, and BD$-$22$^\circ$3467, lie outside the parameter space predicted by previous stable mass-transfer models, even when observational uncertainties are taken into account.
However, all three systems can be reproduced by the stable mass-transfer models with intermediate-mass donors presented in this work.

Therefore, our results suggest that the long-period WD binaries that deviate from the canonical $M_{\mathrm{WD}}$--$P_{\mathrm{orb}}$ relation for low-mass progenitors were most likely formed through stable mass transfer rather than a CEE channel.

\subsubsection{WD+MS Binaies from  Gaia DR3}
The Gaia DR3 catalog has revealed a substantial population of WD+MS binary candidates. Among the WD+MS binaries identified by \citet{shahafTriageGaiaDR32024}, the orbital periods and companion masses have been characterized by \citet{hallakounDeficitMassiveWhite2024}. The observed systems are concentrated around $M_{\mathrm{WD}}\simeq0.6\,M_\odot$, with companion masses between $0.452\,M_\odot$ and $0.858\,M_\odot$.

\begin{figure*}[htbp]
	\centering 
	\includegraphics[width=0.9\textwidth]{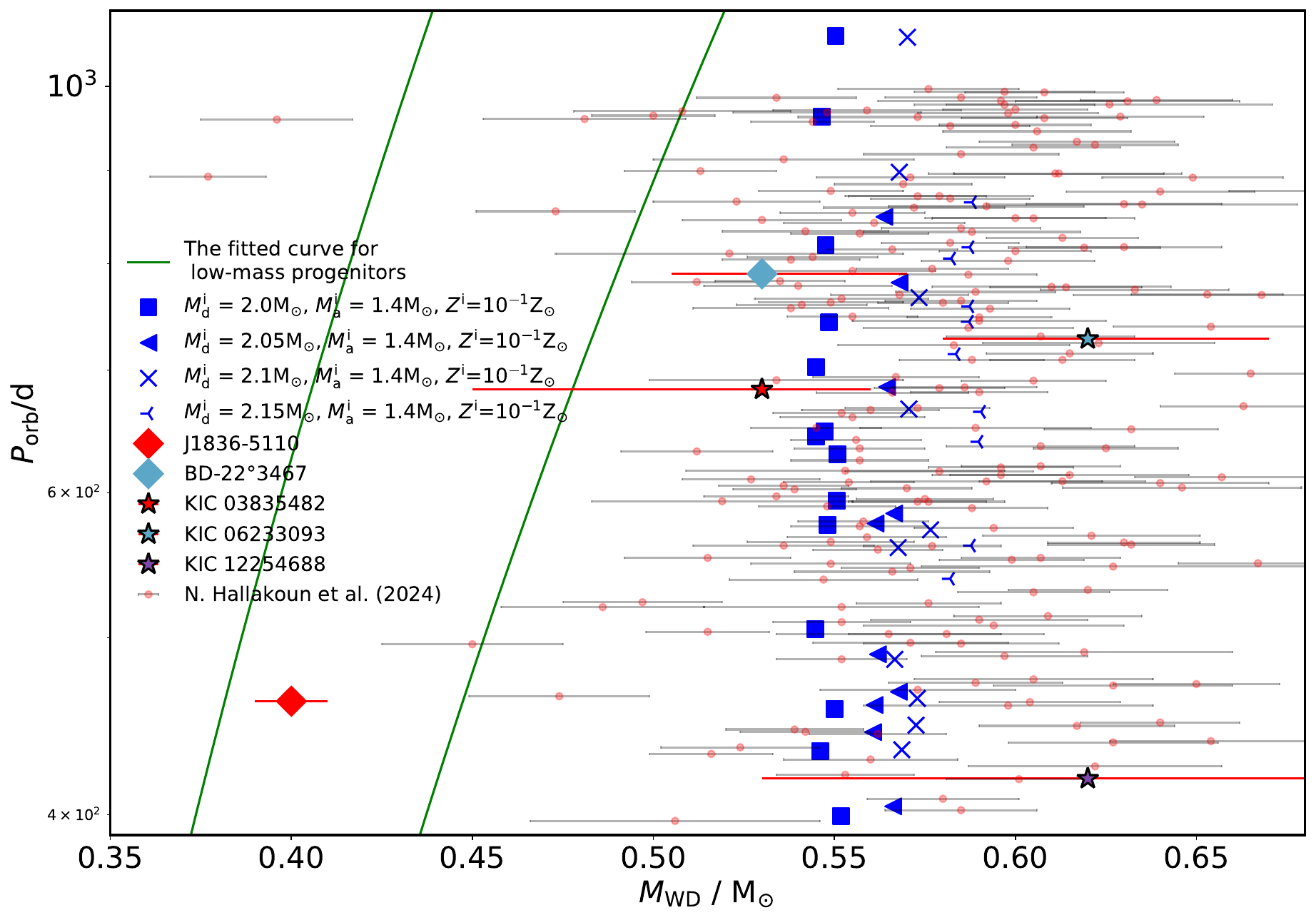} 
\caption{
Comparison between our theoretical $M_{\mathrm{WD}}$--$P_{\mathrm{orb}}$ distributions and observational data. 
The observed systems include Gaia DR3 WD+MS binaries: the orbital periods and masses plotted here are adopted from \citet{hallakounDeficitMassiveWhite2024}, while the systems were originally identified by \citet{shahafTriageGaiaDR32024}. 
Also shown are three self-lensing binaries (KIC 03835482, 06233093, and 12254688) from \citet{kawaharaDiscoveryThreeSelflensing2018}, the WD+subgiant binary J1836--5110 \citep{parsonsWhiteDwarfBinary2022}, and BD$-$22$^\circ$3467 \citep{lobling2020first,bhattacharjee2026thermally}. 
All models assume $\beta = 0.5$, an accretor mass of $1.4\,M_\odot$, and a metallicity of $0.1\,Z_\odot$. 
The green curves in the left and right panels represent the $M_{\mathrm{WD}}$--$P_{\mathrm{orb}}$ relations for low-mass progenitors with $Z_\odot$ and $0.1\,Z_\odot$ metallicities, respectively.
}

	\label{fig-5-c} 
\end{figure*}

As shown in Figure~\ref{fig-5-c}, the majority of the Gaia DR3 WD+MS binaries lie below the canonical $M_{\mathrm{WD}}$--$P_{\mathrm{orb}}$ relation for low-mass progenitors. Within the explored parameter space, the intermediate-mass donor models at $0.1\,Z_\odot$ naturally reproduce this behavior: the resulting $M_{\mathrm{WD}}$--$P_{\mathrm{orb}}$ distribution is shifted toward shorter orbital periods than the canonical relation, even for similar or slightly larger WD masses. As discussed in Section~\ref{sec-Distribution for Progenitors with intermediate mass}, this offset is primarily determined by the core properties of the donor during its evolution, rather than by the initial mass ratio. Therefore, the agreement between our models and the Gaia DR3 systems supports the interpretation that many of these wide WD+MS binaries originated from stable mass transfer involving intermediate-mass progenitors.

However, the companion masses of the Gaia DR3 systems suggest that the progenitor binaries that formed these WD+MS systems likely had larger initial mass ratios than those considered in the present calculations. Therefore, although the agreement is encouraging, a model grid covering larger initial mass ratios is required before firm conclusions can be drawn regarding the origin of these systems.

Previous studies have suggested that some wide WD binary systems may originate from the CEE channel \citep{yamaguchiWidePostcommonEnvelope2023}. However, our results indicate that stable mass transfer from intermediate-mass progenitors is a viable, and possibly necessary, alternative formation pathway.

Overall, the comparison with observations of WD+NS and WD+MS binaries suggests that deviations from the canonical $M_{\mathrm{WD}}$--$P_{\mathrm{orb}}$ distribution can be explained by at least two distinct evolutionary channels: (1) stable mass transfer from extremely metal-poor progenitors and (2) stable mass transfer from intermediate-mass progenitors. These results highlight the importance of considering both metallicity and the internal structure of the donor star in modeling long-period WD binaries.

\section{Discussion} 
\label{sec:discussion}
We note that our results for the period--mass distribution show a close resemblance to the modeled period--luminosity distribution reported by \citet{moltzerUnderstandingPostredGiant2025}, in particular for their simulated post-AGB binaries (see their Fig. 4).

While they adopt luminosity as the primary observable, these two quantities are tightly correlated for post-RGB and post-AGB stars via the well-established core mass--luminosity relation \citep[e.g.,][]{paczynski1970evolution,vassiliadis1993evolution}. Consequently, the relative positioning of their post-AGB sequences in the $P_{\mathrm{orb}}$--$L$ plane is qualitatively consistent with the $M_{\mathrm{WD}}$--$P_{\mathrm{orb}}$ trends shown in our Figures~\ref{fig-3-2-a} and ~\ref{fig-3-2-a0}.

We now turn to a discussion of the physical factors that may influence the predicted $M_{\mathrm{WD}}$--$P_{\mathrm{orb}}$ distribution, as well as the limitations of the present models.

\subsection{Impact of the Accretor}

In numerical simulations, the accretor is often treated as a point mass to simplify the computational cost. 
However, the distinction between an NS accretor and a noncompact accretor (such as a MS star) manifests in several key ways.
(1) Mass-loss Fractions: The fraction of mass lost from the vicinity of the accretor varies, which is typically parameterized by $\beta$ in the MESA code.
(2) Accretion limits: NS accretors are typically limited by the Eddington mass-transfer rate, although super-Eddington accretion may occur in some systems, whereas noncompact accretors generally are not subject to such constraints.
(3) Impacts from radiation: For NS accretors, X-ray radiation during the mass-transfer process can suppress mass-transfer efficiency and carry away additional mass and angular momentum from the system.
These differences collectively influence the binary's mass and angular momentum loss, as dictated by varying mass-transfer efficiencies. 
\begin{figure}[htbp] 
	\centering \includegraphics[width=\columnwidth]{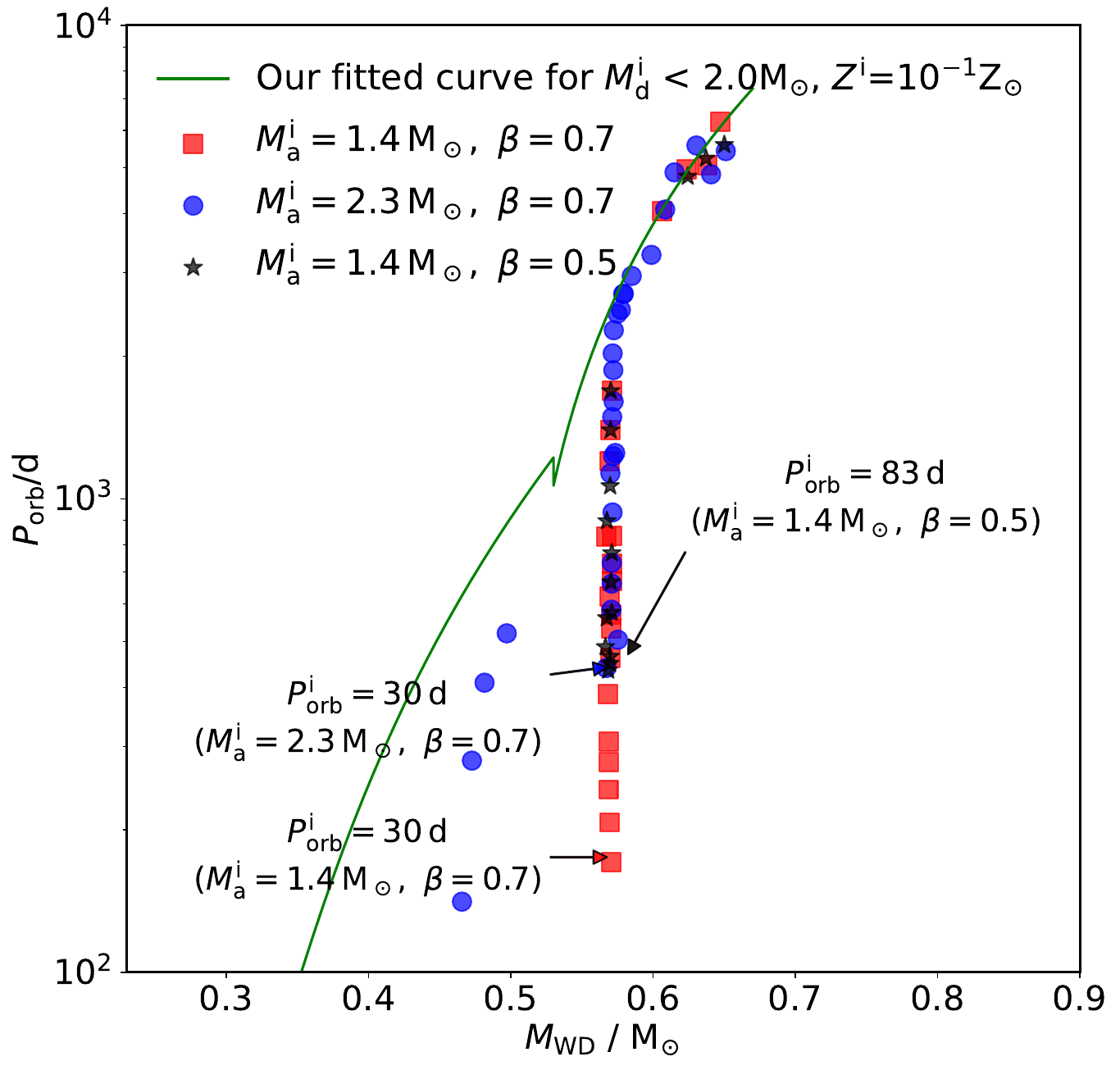} 
	\caption{Distribution of $M_{\mathrm{WD}}$--$P_{\mathrm{orb}}$ for two accretor masses and mass-transfer efficiencies. The models are calculated for an initial donor mass of $2.1\,M_\odot$ and an initial metallicity of $10^{-1}\,Z_\odot$. Arrows indicate the minimum initial orbital periods required for systems to reach a core mass of $0.57\,M_\odot$ while maintaining stable mass transfer.}
	\label{fig:4-2-different accretors} 
\end{figure}

Previous studies have demonstrated that, for low-mass progenitors, the resulting stable $M_{\mathrm{WD}}$--$P_{\mathrm{orb}}$ relation is largely independent of the accretor's mass and mass-transfer efficiency \citep{smedleyNatureMillisecondPulsars2014}. Our results suggest that a similar decoupling holds for the $M_{\mathrm{WD}}$--$P_{\mathrm{orb}}$ distribution produced by intermediate-mass progenitors that deviate from the standard relation. As discussed in Section~\ref{sec-Distribution for Progenitors with intermediate mass}, the "vertical" components in Figure~\ref{fig-3-2-a}—which represent deviations from the typical period relation—originate from cases where the progenitor's core remains nondegenerate before central helium burning.
In these instances, the final $M_{\mathrm{WD}}$--$P_{\mathrm{orb}}$ distribution is primarily determined by the mass of the donor prior to its core becoming degenerate, rather than the specific mass or mass-transfer efficiency of the accretor.

Both the accretor mass and the accretion efficiency significantly affect the stability of mass transfer. As shown in Figure~\ref{fig:4-2-different accretors}, for an accretor with a mass of $2.3\,M_\odot$, mass transfer can occur at shorter initial orbital periods, allowing the resulting $M_{\mathrm{WD}}$--$P_{\mathrm{orb}}$ distribution to extend toward the lower-mass regime.

During binary mass transfer, 
the orbital separation is jointly affected by multiple factors. A smaller amount of mass loss implies a reduced loss of orbital 
angular momentum, which tends to drive a faster increase in the orbital separation. However, 
at the same time, it also implies a slower decrease in the total mass of the system, which tends 
to slow down the increase of the orbital separation. In our models, for the case of $\beta = 0.5$ 
(without considering the Eddington accretion limit and X-ray effects), the orbital separation 
increases relatively slowly during mass-transfer. As a result, the subsequent mass transfer rate 
becomes higher, making the system more prone to instability.

We note that \citet{bhattacharjee2026thermally} found that inflation of the accretor during mass transfer may enhance stability by increasing the effective Eddington accretion rate and therefore the mass-retention efficiency. By contrast, our results indicate that, for the systems considered here, the orbital response to nonconservative mass transfer plays a more important role in determining the stability boundary. These findings highlight that mass-transfer stability is governed by the combined effects of accretor physics, mass-ratio evolution, orbital-separation evolution, and changes in the total binary mass.

For different accretor masses and accretion efficiencies, the resulting WD $M_{\mathrm{WD}}$--$P_{\mathrm{orb}}$ distribution is generally concentrated around the progenitor's core mass at the onset of degeneracy of the donor. At longer orbital periods, the distribution tends to follow the $M_{\mathrm{WD}}$--$P_{\mathrm{orb}}$ relation corresponding to low-mass progenitors. Higher accretor masses and higher mass-loss fractions (i.e., lower $\beta$) both act to stabilize the mass transfer process. The models indicated by blue points in Figure~\ref{fig:4-2-different accretors} illustrate cases where stable mass transfer can occur at shorter orbital periods, leading to a smaller range of resulting WD masses.

For comparison, calculations for the fully conservative case ($\beta = 0$) were also performed. We find that, for systems with an initial accretor mass of $1.4\,M_\odot$ and initial donor masses $M_{\rm d}^{\rm i} > 2.0\,M_\odot$ at metallicities of $Z^{\rm i} = Z_\odot$ and $10^{-1}\,Z_\odot$, the majority of models within the initial period range of 3--1950 days either encounter unstable mass transfer or experience mass loss via the outer Lagrangian points.

\subsection{Model Limitations and Extreme Mass Ratios}

A key physical constraint of our current modeling involves the applicability of the mass-transfer prescription at high mass ratios. However, many of the Gaia DR3 systems discussed in Section~\ref{susubsection obser of WD+MS} involve mass ratios that exceed this limit ($q\,>\,2.0$). In such extreme regimes, the donor's radius may significantly exceed its Roche lobe, leading to mass loss through the outer Lagrangian points $\mathrm{L}_3$. Whether such mass loss inevitably triggers a CEE remains an open and complex question. Notably, our calculations, consistent with the findings of \citet{geAdiabaticMassLoss2020a} and \citet{geThermalEquilibriumMassloss2020}, suggest that systems undergoing substantial mass loss through the $\mathrm{L}_3$ points can remain both dynamically stable and in local thermal equilibrium. This implies that the progenitors of the observed WD binaries systems of Gaia DR3 could potentially survive the extreme mass-transfer phase without entering a CEE. To improve the fidelity of future simulations, it is essential to incorporate modified mass-transfer rates that specifically account for this $\mathrm{L}_3$ overflow. Determining the exact physical boundaries that separate these stable sequences from true CEE events will be critical for a complete understanding of binary evolution at extreme mass ratios.

We anticipate that a stable mass-transfer parameter space persists even when mass loss occurs via the outer Lagrangian points; as a result, the distribution of orbital periods and mass-transfer timescales may be considerably broader than our current results indicate. Determining the precise physical boundaries that separate these stable sequences from true CEE events remains a critical task for future high-fidelity simulations.
\section{Conclusion}
\label{conclusion}

We calculated a large number of binary models\footnote{The inlists used in our calculations and the corresponding data are available in the AAS Journals community on Zenodo at doi: 10.5281/zenodo.21937383.} that formed WDs through stable mass-transfer processes. We employed the quasi-adiabatic criterion to ensure that our models have undergone stable mass-transfer processes. Our initial donor masses span both the low-mass and intermediate-mass regimes. We compare the resulting $M_{\mathrm{WD}}$--$P_{\mathrm{orb}}$ relations obtained under different mass-transfer schemes and initial metallicities. 
We compared our models with the observed long-period WD binary systems considered in this work, including systems such as B0820+02, BD$-$22$^\circ$3467, KIC 03835482, KIC 06233093, and KIC 12254688, several of which deviate from the canonical $M_{\mathrm{WD}}$--$P_{\mathrm{orb}}$ relation predicted by previous studies. Our results suggest that these systems are likely to have formed through stable mass transfer rather than a common-envelope evolution phase.

\begin{enumerate}
	\item For low-mass progenitors, we provide formulae that can fit the $M_{ \mathrm{WD}}$--$P_{ \mathrm{orb}}$ relations resulting from different mass-transfer schemes and various initial metallicities.
    
	\item The $M_{ \mathrm{WD}}$--$P_{ \mathrm{orb}}$ relation is significantly influenced by the initial metallicity and the core properties of the progenitor. As the metallicity decreases, the $M_{ \mathrm{WD}}$--$P_{ \mathrm{orb}}$ relation shifts downward \citep{rappaportRelationWhiteDwarf1995}.
    For intermediate-mass progenitors with nondegenerate cores, the resulting $M_{ \mathrm{WD}}$--$P_{ \mathrm{orb}}$ relation is lowered in the short-period regime, while in the long-period regime it approaches the relation produced by low-mass progenitors.
	
	\item Compared with the Kolb scheme, the Han scheme is more prone to driving mass transfer into an unstable regime. In addition, models with solar metallicity exhibit a larger parameter space for unstable mass transfer than those with one-tenth solar metallicity. 
    For extremely low-metallicity models, donors with intermediate initial masses exhibit a larger parameter space for stable mass transfer.
    This indicates that both the adopted mass-transfer prescription and the metallicity play important roles in shaping the distribution of the $M_{\mathrm{WD}}$--$P_{\mathrm{orb}}$ parameter space.

   \item By comparing our theoretical models with a diverse sample of WD+NS and WD+MS binaries, we demonstrate that the observed departures from the canonical $M_{\mathrm{WD}}$--$P_{\mathrm{orb}}$ relation (i.e. the theoretical relations for low-mass progenitors at solar and $0.1\,Z_\odot$ metallicity) are primarily driven by two distinct evolutionary channels: (i) stable mass transfer from low-mass, metal-poor progenitors, and (ii) stable mass transfer from intermediate-mass progenitors where substantial core growth occurs before the onset of strong electron degeneracy. Specifically, our models successfully reproduce the parameters of long-period self-lensing binaries (e.g., KIC 06233093 and KIC 12254688), BD$-$22$^\circ$3467 and the pulsar binary B0820+02, which lie within the expanded $M_{\mathrm{WD}}$--$P_{\mathrm{orb}}$ parameter space predicted in this work. These results suggest that stable mass transfer is a more prevalent formation channel for long-period WD binaries than previously assumed, in particular for systems located below the standard relation. Our findings highlight the necessity of accounting for the donor's internal structure and metallicity-dependent evolution to accurately interpret the formation history of the growing population of observed WD systems.

    \item For the WD binaries identified in Gaia DR3, although our results are able to cover their $M_{\mathrm{WD}}$--$P_{\mathrm{orb}}$ parameter space, we cannot firmly conclude that they were formed through stable mass transfer. Considering the extreme mass ratios of their progenitor systems, it is necessary to further revise the mass-transfer rates and the criterion for the onset of a CEE triggered by mass loss through the outer Lagrangian points. Such improvements are essential for assessing whether these systems can indeed be produced via stable mass transfer.

\end{enumerate}

\begin{acknowledgements}
This project is supported by the Strategic Priority Research Program of the Chinese Academy of Sciences (grant No. XDB1160201), the National Natural Science Foundation of China (NSFC Nos. 12288102, 12525304, 12125303, 12473033, 12333008, 12233013, and 11803030), the National Key R\&D Program of China (No. 2021YFA1600403), Yunnan Revitalization Talent Support Program—Science \& Technology Champion Project (No. 202305AB350003), the Yunnan Fundamental Research Projects (No. 202401BC070007), the International Centre of Supernovae, Yunnan Key Laboratory (No. 202302AN360001), CAS Project for Young Scientists in Basic Research (YSBR-148), the Yunnan Revitalization Talent Support Program "YunLing Scholar" project, and the New Cornerstone Science Foundation through the XPLORER PRIZE. This work is partially supported by the National Natural Science Foundation of China (grant Nos: 12333008 and 12422305), the Joint Research Fund in Astronomy U2031203. C.A.T. thanks Churchill College for his Fellowship. We gratefully acknowledge the 'PHOENIX Supercomputing Platform" jointly operated by the Binary Population Synthesis Group and the Stellar Astrophysics Group at Yunnan Observatories, Chinese Academy of Sciences. We are grateful to the anonymous reviewer for the insightful remarks and constructive feedback, which significantly improved the clarity and quality of this work.

\end{acknowledgements}

\bibliography{sample7}{}

\bibliographystyle{aasjournalv7}
\end{document}